\documentclass[twocolumn,trackchanges]{aastex701}
\usepackage{amsmath}

\begin{document}

\title[Formation of ELM WD with EAML]{Formation of extremely low-mass white dwarf binaries undergoing enhanced angular momentum loss}

\author[orcid=0009-0002-0904-3720]{Ziqi Zhao}
\affiliation{Yunnan Observatories, Chinese Academy of Sciences (CAS),
Kunming 650216, People’s Republic of China}
\affiliation{International Centre of Supernovae (ICESUN), Yunnan Key Laboratory of Supernova Research, Kunming 650216, People’s Republic of China}
\affiliation{University of Chinese Academy of Sciences, Beijing 100049, People’s Republic of China}
\email{zhaoziqi@ynao.ac.cn}

\author[orcid=0000-0002-1421-4427]{Zhenwei Li}
\affiliation{Yunnan Observatories, Chinese Academy of Sciences (CAS),
Kunming 650216, People’s Republic of China}
\affiliation{International Centre of Supernovae (ICESUN), Yunnan Key Laboratory of Supernova Research, Kunming 650216, People’s Republic of China}
\affiliation{University of Chinese Academy of Sciences, Beijing 100049, People’s Republic of China}
\email{lizw@ynao.ac.cn}

\author[orcid=0000-0002-7909-4171]{Zhengwei Liu}
\affiliation{Yunnan Observatories, Chinese Academy of Sciences (CAS),
Kunming 650216, People’s Republic of China}
\affiliation{International Centre of Supernovae (ICESUN), Yunnan Key Laboratory of Supernova Research, Kunming 650216, People’s Republic of China}
\affiliation{University of Chinese Academy of Sciences, Beijing 100049, People’s Republic of China}
\email{zwliu@ynao.ac.cn}

\author[orcid=0009-0006-9211-2860]{Hailiang Chen}
\affiliation{Yunnan Observatories, Chinese Academy of Sciences (CAS),
Kunming 650216, People’s Republic of China}
\affiliation{International Centre of Supernovae (ICESUN), Yunnan Key Laboratory of Supernova Research, Kunming 650216, People’s Republic of China}
\affiliation{University of Chinese Academy of Sciences, Beijing 100049, People’s Republic of China}
\email{chenhl@ynao.ac.cn}

\author[orcid=0000-0002-6398-0195]{Hongwei Ge}
\affiliation{Yunnan Observatories, Chinese Academy of Sciences (CAS),
Kunming 650216, People’s Republic of China}
\affiliation{International Centre of Supernovae (ICESUN), Yunnan Key Laboratory of Supernova Research, Kunming 650216, People’s Republic of China}
\affiliation{University of Chinese Academy of Sciences, Beijing 100049, People’s Republic of China}
\email{gehw@ynao.ac.cn}

\author[orcid=0000-0001-5316-2298]{Xiangcun Meng}
\affiliation{Yunnan Observatories, Chinese Academy of Sciences (CAS),
Kunming 650216, People’s Republic of China}
\affiliation{International Centre of Supernovae (ICESUN), Yunnan Key Laboratory of Supernova Research, Kunming 650216, People’s Republic of China}
\affiliation{University of Chinese Academy of Sciences, Beijing 100049, People’s Republic of China}
\email{xiangcunmeng@ynao.ac.cn}

\author[orcid=0000-0003-4265-7783]{Dengkai Jiang}
\affiliation{Yunnan Observatories, Chinese Academy of Sciences (CAS),
Kunming 650216, People’s Republic of China}
\affiliation{International Centre of Supernovae (ICESUN), Yunnan Key Laboratory of Supernova Research, Kunming 650216, People’s Republic of China}
\affiliation{University of Chinese Academy of Sciences, Beijing 100049, People’s Republic of China}
\email{dengkai@ynao.ac.cn}

\author[orcid=0000-0001-5284-8001]{Xuefei Chen}
\affiliation{Yunnan Observatories, Chinese Academy of Sciences (CAS),
Kunming 650216, People’s Republic of China}
\affiliation{International Centre of Supernovae (ICESUN), Yunnan Key Laboratory of Supernova Research, Kunming 650216, People’s Republic of China}
\affiliation{University of Chinese Academy of Sciences, Beijing 100049, People’s Republic of China}
\email{cxf@ynao.ac.cn}

\author[orcid=0000-0001-9204-7778]{Zhanwen Han}
\affiliation{Yunnan Observatories, Chinese Academy of Sciences (CAS),
Kunming 650216, People’s Republic of China}
\affiliation{International Centre of Supernovae (ICESUN), Yunnan Key Laboratory of Supernova Research, Kunming 650216, People’s Republic of China}
\affiliation{University of Chinese Academy of Sciences, Beijing 100049, People’s Republic of China}
\email{zhanwenhan@ynao.ac.cn}

\correspondingauthor{Zhenwei Li, Zhengwei Liu, Xuefei Chen}
\email{lizw@ynao.ac.cn, zwliu@ynao.ac.cn, cxf@ynao.ac.cn}

%\collaboration{all}{The Terra Mater collaboration}

\begin{abstract}
Extremely low-mass white dwarfs (ELM WDs) are helium (He) WDs with masses below $\sim 0.3\ M_{\odot}$, mainly formed through binary interaction. ELM WD binaries typically are formed from two channels, namely the stable Roche lobe overflow (RLOF) channel and the common envelope ejection channel. For ELM WD binaries produced from RLOF channel, the ELM WD mass has a strong correlation with the orbital period, i.e., the so-called WD mass-orbital period relation. 
However, the observations in the ELM Survey show that the orbital periods of ELM WD binaries from the RLOF channel are typically shorter than the theoretically predicted values. Extra angular momentum loss (AML) may be needed to explain such a phenomenon. In this work, we assumed that part of the transferred mass from the donor is lost at the outer Lagrangian point and simulated the formation of ELM WD binaries. Enhanced AML enables more mass to be lost during thermal-timescale mass transfer, thereby affecting nuclear burning in the transfer phase and producing ELM WDs with distinct internal structures. These structural differences alter the (pre-)He WD mass–radius relation at the end of mass transfer, which in turn shifts the WD mass–orbital period relation downward. These adjustments enable our model to successfully reproduce the majority of observed systems from the relevant survey projects. 
\end{abstract}

\keywords{\uat{Binary stars}{154} --- \uat{Close binary stars}{254} --- \uat{White dwarf stars}{1799}}


\section{Introduction}
\label{intro}

Extremely low-mass white dwarfs (ELM WDs), defined as helium white dwarfs (He WDs) with masses below approximately $0.3\ M_{\odot}$ \citep{2016ApJ...824...46B}. Due to their extremely low mass, it is widely accepted that nearly all known ELM WDs originate from binary evolution rather than through single-star evolution \citep[e.g.,][]{1990ApJ...353..215I, 1995MNRAS.275..828M, 2020ApJ...889...49B}. This characteristic renders them particularly valuable for investigating stellar physics and binary interactions. Besides, many ELM WD binaries are found in close orbits, which makes them become one of the most important gravitational wave (GW) sources
\citep[e.g.,][]{2017arXiv170200786A,Li_2025}.

ELM WDs have been identified by several survey projects, such as the Kepler project \citep{2010ApJ...715...51V,2011ApJ...728..139C,2012ApJ...748..115B,2015ApJ...803...82R,2015ApJ...815...26F,2017ApJ...851...39G,2019ApJ...884..165Z}, the ELM Survey \citep{2010ApJ...723.1072B,2011ApJ...727....3K,2012ApJ...744..142B,2012ApJ...751..141K,2013ApJ...769...66B,2015ApJ...812..167G,2016ApJ...818..155B,2020ApJ...889...49B,2020ApJ...894...53K, 2022ApJ...933...94B,2023ApJ...950..141K}, the Wide Angle Search for Planets \citep{2011MNRAS.418.1156M,2014MNRAS.444..208M,2020NewA...7801363L,2020AJ....160...49L,2021AJ....161..137H,2022MNRAS.511..654L,2024ApJ...973..114L}, the LAMOST\citep{2022ApJ...933..193Z,2022ApJ...936....5W,2023AJ....165..119Y,2023MNRAS.526.5471Y} and AstroSat's Ultra Violet Imaging Telescope (UVIT) \citep{2023ApJ...943..130D,2024BSRSL..93..300D}. In particular, the ELM Survey has discovered over 100 detached double degenerate binaries with ELM WDs, providing the largest and most well-characterized sample for studying their formation and evolution. 

The key to forming an ELM WD is that the progenitor needs to strip the H-rich envelope as it develops the He core.  Two formation channels are introduced, i,.e., stable Roche lobe overﬂow (RLOF) channel \citep{2013A&A...557A..19A, 2016A&A...595A..35I, 2017MNRAS.467.1874C, 2018ApJ...858...14S, 2019ApJ...871..148L} and  common envelop (CE) channel \citep{2016PASP..128h2001H, 2019ApJ...871..148L, 2025A&A...699A.280A}. For ELM WDs from RLOF channel, the ﬁnal ELM WD mass and the orbital period follow a tight relation, which is known as WD mass-period relation \citep[e.g.,][]{1995MNRAS.273..731R, 1999A&A...350..928T, 2004ApJ...616.1124N, 2011ApJ...732...70L, 2014A&A...571A..45I, 2014ApJ...791..127J,2017MNRAS.467.1874C, 2021MNRAS.503.3540C,2025A&A...700A.219N}. For ELM WDs from the CE channel, the rapid orbital shrinkage during the CE phase generally leads to a close orbit of the ELM WD binary. Therefore, 
ELM WD binaries produced from these two channels should show distinctive observed characteristics \citep{2018A&A...614A..49C}. Indeed, observations show that the ELM WD binaries can be divided into two groups, which are consistent with the theoretical predictions \citep{2019ApJ...871..148L, 2020ApJ...889...49B}. 

Although theoretical models and observational data are largely consistent, discrepancies remain in certain details. Specifically, for ELM WDs formed through the stable RLOF channel, the observed orbital periods in the ELM Survey are typically shorter than theoretically predicted \citep{2019ApJ...871..148L}. For such binaries, the final orbital period is primarily determined by two factors: the stellar structure of the donor and the rate of angular momentum loss (AML). In a recent study, \citet{2025arXiv251120147M} adopted opacity tables based on \citet{2008ApJS..174..504F,2014ApJS..214...25F} and found that donors evolve faster and start mass transfer earlier, resulting in shorter orbital periods for a given ELM WD mass. In their studies, most of the observations can be explained by the models with metallicity Z = 0.02 and 0.001.
But models with a metallicity of Z = 0.0001 to explain several systems.
In this work, we investigate the role of enhanced AML in the formation of ELM WD binaries. 
Enhanced AML alters orbital evolution, and for a given final ELM WD mass, donors may exhibit different internal structures depending on the strength of AML.
Therefore, exploring enhanced AML offers a promising alternative pathway to resolve the period discrepancy without invoking unrealistic metallicity assumptions.

The structure of this paper is as follows. Section 
\ref{Formation channels of ELM WD binaries} re-classifies the formation channels based on the latest ELM survey samples. The model inputs and methods are described in Section \ref{sect:Model Inputs and Methods}, and the results are shown in Section \ref{results}. Section \ref{Mass loss versus MB} discusses the distinction between AML via magnetic braking (MB) and via mass loss. Finally, the conclusions are summarized in Section \ref{conclusion}.

\section{Formation channels of ELM WD binaries}
\label{Formation channels of ELM WD binaries}

In this work, we primarily focus on the ELM WDs with CO WD companions,
which are the most common binary type observed in the ELM Survey. The formation of such systems requires two distinct phases of mass transfer. In the first phase, the primordial binary undergoes a CE ejection, leaving behind a binary composed of a CO WD and a main-sequence (MS) star.
Subsequently,
the MS star evolves and fills its Roche lobe,
initiating the second mass transfer phase.
This phase may proceed either stably (RL channel) or unstably (CE channel), producing ELM WD binaries with distinct physical characteristics \citep{2019ApJ...871..148L}.

Adopting a similar methodology to \citet{2019ApJ...871..148L}, we re-classify the formation channels for the updated observed samples of ELM WD binaries in the ELM survey \citep{2016ApJ...818..155B, 2020ApJ...889...49B, 2022ApJ...933...94B}. We assume that all the ELM WD samples are formed from CE ejection. Then for each sample, we can calculate a value of CE efficiency, $\alpha_{\rm CE}$, based on the standard energy budget equation \citep{1984ApJ...277..355W,1988ApJ...329..764L}:
\begin{equation}
\label{eq12}
\alpha_{\rm{CE}}=\frac{E_{\mathrm{bind}}}{G\left(M_{\text {c }}+M_{\text {env }}\right) M_{\mathrm{a}} /\left(2 a_{\mathrm{i}}\right)-G M_{\text {c }} M_{\mathrm{a}} /\left(2 a_{\mathrm{f}}\right)},
\end{equation}
where $M_{\rm c}$ and $M_{\rm env}$ are the donor's core and envelope masses, respectively. $G$ is the gravitational constant, and $M_{\rm a}$ is the accretor mass, taken as the CO WD mass from observations in this work. The final binary separations $a_{\rm{f}}$, can be derived from observations, and the initial separation, $a_{\rm{i}}$, can be estimated as \citep{1983ApJ...268..368E}: 
\begin{equation}
a_{\mathrm{i}}=R_{\mathrm{RL}} \frac{0.6 q^{2 / 3}+\ln \left(1+q^{1 / 3}\right)}{0.49 q^{2 / 3}},
\end{equation}
where $R_{\rm{RL}}$ is the Roche lobe radius, and $q = M_{\rm d}/M_{\rm a}$ is the mass ratio, with $M_{\rm d}$ denoting the donor mass.

The envelope binding energy, $E_{\mathrm{bind}}$, is generally expressed as  \citep{1995MNRAS.272..800H, 2000A&A...360.1043D, 2013A&ARv..21...59I, 2020RAA....20..161H,2024PrPNP.13404083C}
\begin{equation}
E_{\mathrm{bind}}= E_{\rm{gr}}+\alpha_{\rm{th}} E_{\rm{th}}, 
\end{equation}
where $E_{\rm gr} = \int_{\rm core}^{\rm surface} -\frac{Gm}{r}\,{\rm d}m$ is the gravitational binding energy of the envelope.
The thermal energy is given by $E_{\rm th} = \int_{\rm core}^{\rm surface} U\,{\rm d}m$, 
where $U$ is the specific internal energy of the gas, including both thermal and recombination energy \citep[e.g.,][]{ 2011ApJS..192....3P, 2015MNRAS.447.2181I, 2021A&A...645A..54K}.
The parameter $\alpha_{\rm th}$ (typically $0-1$) represents the contribution of thermal energy to the CE ejection.
We adopt $\alpha_{\rm{th}} = 1$ in our simulations,
which means that all the thermal energy contributes to CE ejection.

\begin{figure*}
  \begin{minipage}[t]{1\linewidth}
  \centering
  \includegraphics[scale=0.45]{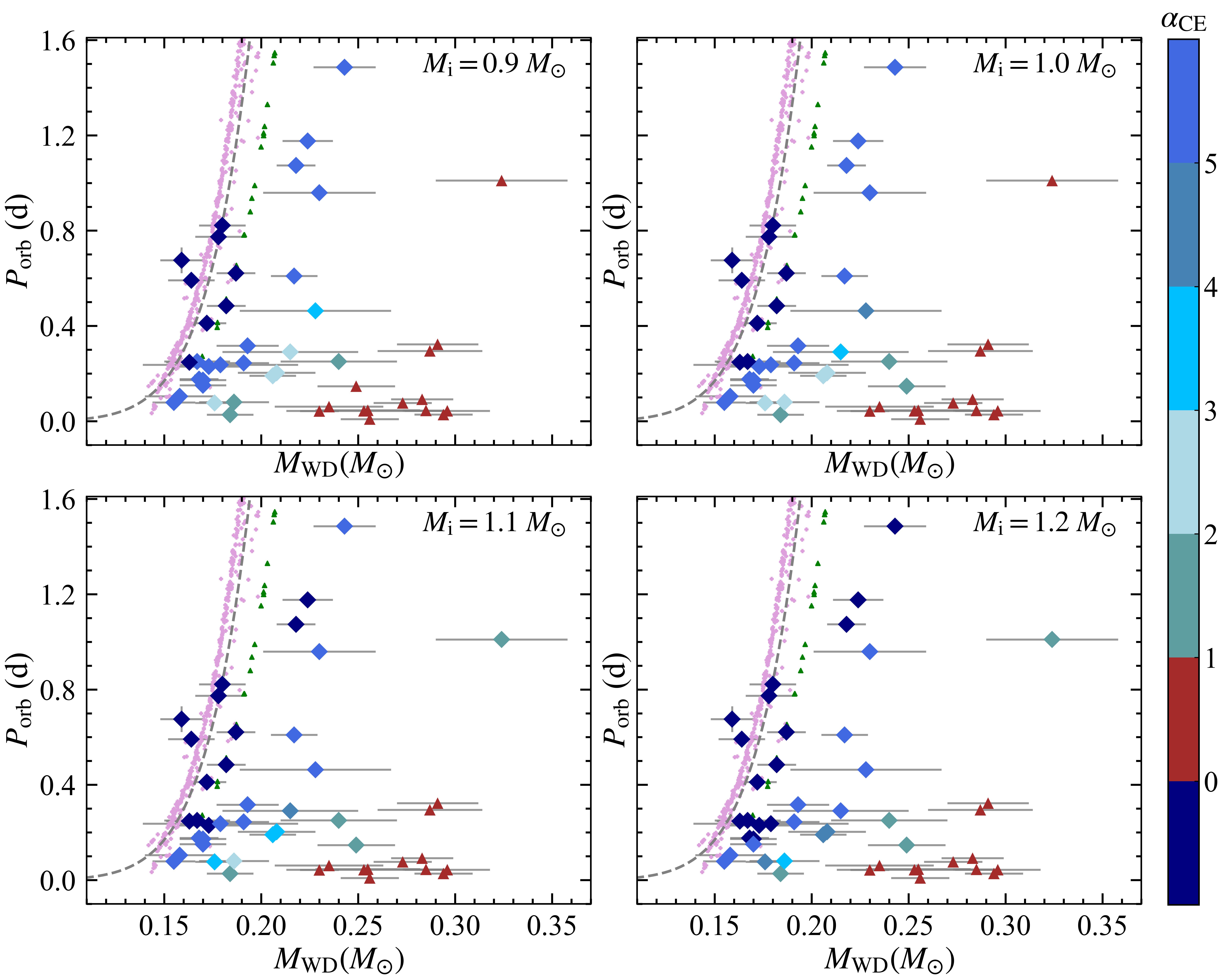}
  \end{minipage}%
\caption{\label{fig1}
The $M_{\rm WD}$–$P_{\rm orb}$ relation for MS progenitors with initial masses of $0.9$–$1.2\ M_{\odot}$. The
fitted ELM WD mass–period relation from \citet{2011ApJ...732...70L} are shown as gray dashed lines. 
The pink crosses and green triangles represent the $Z=0.02$ and $Z=0.001$ models from \citet{2019ApJ...871..148L}, respectively. 
Observed ELM WD systems are taken from \citet{2016ApJ...818..155B} and updated in \citet{2020ApJ...889...49B,2022ApJ...933...94B}. 
Assuming that all observed systems formed via CE ejection, we compute the corresponding CE efficiency, $\alpha_{\rm CE}$, as indicated by the colorbar. 
Systems with unreasonable values ($\alpha_{\rm CE}<0$ or $\alpha_{\rm CE}>1$) are instead interpreted as being formed through the RLOF channel and are shown as diamonds, while red triangles denote systems formed via CE ejection.
 }
\end{figure*}

Our calculation starts with the evolution of a single star. Once the core reaches a mass corresponding to an observed ELM WD, we assume the onset of the CE phase. For each such case, we record the binding energy, core mass, and stellar radius, which allows us to subsequently derive the CE efficiency for each observed sample using the aforementioned methodology. The boundary between the core and the envelope is defined at the location where the H mass fraction ($X_{\rm H}$) reaches 0.01.
We also test an alternative definition using $X_{\rm H}=0.1$ as the boundary (see Figure~\ref{fig8} in the Appendix~\ref{appendix}),
and the results show that the choice of the core–envelope boundary criterion does not significantly affect the classification of ELM WD formation channels.

Figure~\ref{fig1} presents the observed samples in the ELM WD mass–orbital period plane, with their corresponding $\alpha_{\rm CE}$ values shown for progenitor masses of 0.9, 1.0, 1.1, and 1.2 $M_\odot$, which represent the typical mass range for ELM WD formation from the CE channel \citep{2019ApJ...871..148L}. The grey dashed line indicates the fitted ELM WD mass–period relation from \citet{2011ApJ...732...70L}, while the pink crosses and green triangles represent the simulated results for metallicities $Z = 0.02$ and $Z = 0.001$, respectively, from \citet{2019ApJ...871..148L}.

The absolute value of the binding energy of the stellar envelope increases with progenitor mass. Hence, a higher $\alpha_{\rm CE}$ is required in order to eject it during the CE phase. 
In an extreme case of 0.9 $M_\odot$ progenitor, many binaries have $\alpha_{\rm CE}>1$ and $\alpha_{\rm CE}<0$, as shown by the colored diamonds in Figure~\ref{fig1}.  These values of $\alpha_{\rm }$ are unreasonable. For the case of $\alpha_{\rm CE}>1$, it means that even if all of the orbital energy is used to eject the envelope, the envelope still cannot be ejected. In particular, some studies suggest typical $\alpha_{\rm CE}$ values are only $\sim 0.3$ for low-mass post-CE systems \citep[e.g.,][]{2010A&A...520A..86Z,2013A&A...557A..87T,2022MNRAS.513.3587Z,2023MNRAS.518.3966S,2024ApJ...977...24Z}. For the case of $\alpha_{\rm CE}<0$, it means that the absolute value of the final orbital energy is larger than that of the initial orbital energy, which is an unreasonable condition in the standard energy budget equation. As a result, those binaries with unreasonable $\alpha_{\rm CE}$ values suggest that they may not produced from the CE evolution but the RL channel. The fact that $\alpha_{\rm CE} > 1$ has previously motivated the development of the $\gamma$-formalism \citep{2000A&A...360.1011N, 2005MNRAS.356..753N, 2025A&A...700A.219N}. However, the $\gamma$-mechanism was originally designed for the first mass transfer phase (evolved star + MS star), whereas our work focuses on the second mass transfer phase (low-mass donor + CO WD), where this mechanism may not be applicable (\citealt{2000A&A...360.1011N}). Nevertheless, we have also attempted to reconstruct the observed sample using the $\gamma$-formalism (see Appendix~\ref{appendixb}) and find that it does not alter our main conclusions.

As shown in Figure~\ref{fig1}, based on the methodology in distinguishing the formation channel of ELM binaries, we can see that $\sim68\%$ ELM WD binaries are formed from the RL channel rather than the CE channel. However, it is clear that the theoretical simulations in the previous studies \citep[e.g.,][]{2011ApJ...732...70L,2019ApJ...871..148L} cannot reproduce the observed ELM WD mass-orbital period distributions. They predict systematically longer orbital periods than observed, as shown in the dashed lines, pink crosses, and green triangles. 

In this work, we focus on ELM binaries produced from the RL channel.
For the observed samples, we conservatively adopt $\alpha_{\rm CE}=1$ as the boundary condition to distinguish between the CE and RL formation channels in the case of a 0.9 $M_\odot$ progenitor.

\section{Model Inputs and Methods}
\label{sect:Model Inputs and Methods}

For a star with a degenerate core, there exists a fairly tight relation between the final orbital period and the WD mass, which is almost independent of AML. In contrast, systems with donors possessing non-degenerate cores deviate from the core mass–radius relation for giants \citep[e.g.,][]{1999A&A...350..928T,2004ApJ...616.1124N,2017MNRAS.467.1874C,2023ApJS..265...15J}. In this work,
we mainly focus on the evolution of WD binaries with donor masses in the range of $1.6-2.8 M_\odot$, in which the donor can develop either a partially degenerate core or a non-degenerate core.
For comparison, the system with a $1.4 M_\odot$ donor that forms a degenerate core is also simulated.

We simulate the detailed binary evolution with the state-of-the-art stellar evolution code Modules for Experiments in Stellar Astrophysics (MESA, version 12115; \citealt{2011ApJS..192....3P, 2013ApJS..208....4P, 2015ApJS..220...15P, 2018ApJS..234...34P, 2019ApJS..243...10P,2023ApJS..265...15J}). Our simulations begin with a CO WD + MS binary system. The initial donor mass $M_{\rm{d,i}}$ is varied from $1.4 M_\odot$ to $2.8 M_\odot$. We test the impact of two different accreted star masses, 0.8 and 1.1 $M_\odot$, on the results.
In addition, the observed ELM WDs include both Galactic halo and disk populations \citep{2016ApJ...824...46B, 2020ApJ...889...49B, 2022ApJ...933...94B}. Accordingly, we consider two metallicities for the donor, i.e., solar metallicity ($Z = 0.02$) and low metallicity ($Z = 0.001$), in order to explore their effects on the binary evolution.

We perform a grid of calculations with different initial orbital periods, with the maximum evolutionary age of our models set to the age of the Universe ($13.7\ \rm{Gyr}$). The upper limits correspond to the orbital periods at which the binary system experiences unstable mass transfer
when the mass transfer rate exceeds $10^{-4}\ M_{\odot}\ {\rm yr}^{-1}$. 
The minimum orbital period corresponds to the formation of a detached WD binary.
We treat overshooting as a diffusive process with an exponentially decaying diffusion coefficient \citep[e.g.,][]{2011ApJS..192....3P,2000A&A...360..952H}.
The extent of the overshoot region is controlled by the free parameter $f_{\rm ov}$,
for which we adopt the recommended value of $0.016$
\citep[e.g.,][]{1992A&AS...96..269S,2000A&A...360..952H, 2025A&A...699A.280A}. Elemental diffusion is not included to simplify the modeling.
In the following, we present the detailed mechanisms of mass transfer and AML.

\subsection{Mass Transfer and Accretion}
\label{Mass Transfer and Accretion}

In this work, we adopt the Kolb scheme to calculate the mass transfer rate through RLOF \citep{1990A&A...236..385K}. 
During the mass transfer phase, the CO WD does not always accumulate the accreted material on its surface but may undergo stable or unstable thermonuclear burning depending on the mass transfer rate and the H abundance of the transferred material\citep{1996ApJ...470L..97H,2004MNRAS.350.1301H,2009MNRAS.395.2103M,2019MNRAS.490.1678C,2024ResPh..5907568L}. 
We define the mass growth rate of the CO WD as
\begin{equation}
\dot{M}_{\text{acc}} = \eta_{\text{H}}\eta_{\text{He}} |\dot{M}_{\text{d}}|,
\end{equation}
where $\eta_{\rm H}$ is the mass accumulation efficiency for H shell burning, which is expressed as 
\begin{equation}
\eta_{\mathrm{H}}=\left\{\begin{array}{l}
\dot{M}_{\mathrm{cr}} /\left|\dot{M}_{\mathrm{d}}\right|, \quad\left|\dot{M}_{\mathrm{d}}\right|>\dot{M}_{\mathrm{cr}} \\
1, \quad \dot{M}_{\mathrm{cr}} \geqslant\left|\dot{M}_{\mathrm{d}}\right| \geqslant \frac{1}{8} \dot{M}_{\mathrm{cr}} \\
0, \quad\left|\dot{M}_{\mathrm{d}}\right|<\frac{1}{8} \dot{M}_{\mathrm{cr}}
\end{array}\right.
\end{equation}
where $\dot{M}_{\text{d}}$ is the mass loss rate from the donor, and $\dot{M}_{\rm cr}$ is the critical accretion rate of the CO WD, with expression of 
\begin{equation}
\label{eq2}
\dot{M}_{\text{cr}} = 5.3 \times 10^{-7}\,M_\odot\,{\rm yr^{-1}}\frac{1.7 -X}{X} ( M_{\text{CO}} - 0.4), 
\end{equation}
The He mass accumulation efficiency, $\eta_{\rm He}$, depends on the CO WD mass $M_{\rm CO}$ and the mass transfer rate, and is adopted from \citet{2004ApJ...613L.129K}.

\subsection{Angular Momentum Loss}
\label{Angular Momentum Loss}

In this paper, we consider three AML mechanisms: mass loss, MB, and GW radiation.
For the case of AML due to the mass loss, many previous works assumed that the non-accreted material would carry away the specific angular momentum of the surface of the accretor \citep{2017MNRAS.471.4256V}. However, the mass transfer process could be more complex in reality \citep{2012ApJ...745..165S,2023MNRAS.525.2605G}. 
For accretion onto a WD, the fate of the transferred material depends on the mass transfer rate relative to the critical rate $\dot{M}{\rm cr}$. If the mass transfer rate exceeds $\dot{M}{\rm cr}$, steady nuclear burning on the WD surface cannot match the accretion rate, leading to the accumulation of material in an extended, red giant-like envelope. When this envelope expands sufficiently to fill its Roche lobe, mass can be lost through the outer Lagrangian point $L_2$ \citep{2017MNRAS.469.4763M}. Conversely, at low mass transfer rates (typically $\lesssim \dot{M}_{\rm cr}/8$), nuclear burning becomes unstable, resulting in nova eruptions once a critical ignition mass is reached \citep{2015ApJ...805L...6S,2024ApJ...977...34T}. Such eruptions may lead to enhanced frictional AML, similar to nova eruptions in cataclysmic variables \citep{2016MNRAS.455L..16S}. According to \citet{2022ApJ...938...31S}, in most nova events the ejecta expand into the circumbinary environment and are likely to escape through the $L_2$ point.

In our models, the mass transfer rates span a wide range. To capture the qualitative effect of additional AML in a simplified manner, we assume that a fraction of the non-accreted material is lost through the $L_2$ point\footnote{Disk instability may drive $L_2$ outflow in binaries with high mass transfer rates \citep[typically $\gtrsim 10^{-4}\ M_{\odot}\ \mathrm{yr^{-1}}$; see][]{2023MNRAS.519.1409L,2025ApJ...990..172S,2025arXiv251024127S}. Our situation, however, differs from such disk-instability-driven $L_2$ outflows, as the mass transfer rates in our WD binary models are well below this threshold.}
, regardless of the specific mechanism.
Following \citet{2024A&A...681A..31P}, the corresponding change in orbital angular momentum is given by 
\begin{equation}
\label{eq9}
\begin{aligned}
\dot{J}_{\text{ML}} &=  
-(1-\eta)\dot{M}_{\text{d}}\Bigg[{k (x_{\rm{L_2}} - x_{\rm{cm}})^2} \\
&\quad+(1-k)
\left(\frac{M_{\text{d}}}{M_{\text{d}}+M_{\text{CO}}}\right)^{2}\Bigg]
\frac{2\pi a^2}{P_{\text{orb}}},
\end{aligned}
\end{equation}
where $a$ is the binary separation, $P_{\rm orb}$ is the orbital period, and $\eta\equiv\eta_{\rm H}\eta_{\rm He}$ denotes the total accumulation efficiency. The parameter $k$ represents the fraction of AML carried away through $L_2$.
Although the parameter $k$ may vary during the real mass transfer process \citep{2024ApJ...977...34T}, for simplicity, it is taken to be constant in this work.
We varied the parameter $k$ from 0 to 0.25 in steps of 0.05.
A larger value of $k$ makes the binary system more prone to unstable mass transfer.
We test the case with $k = 0.3$
and find that none of the binary systems can undergo stable RLOF.
Moreover, adopting $k = 0.25$ as the upper limit is sufficient to account for the majority of the observed systems.
$k=0$ represents the standard AML, which is commonly assumed in many previous works \citep[e.g.,][]{2016A&A...595A..35I,2018ApJ...858...14S,2019ApJ...871..148L,2025arXiv251120147M}. In our simulations, the orbit is assumed to be circular.

The distance between the outer Lagrangian point, $x_{\rm L_2}$, and the center of mass $x_{\rm cm}=1/(1+q)$ of the system, 
is described by the separation $x_{\rm L_2}-x_{\rm cm}$. \citet{2024A&A...681A..31P} calculated a numerical solution for $(x_{\rm L_2}-x_{\rm cm})^2$, which is
\begin{equation}
(x_{\rm{L_2}} - x_{\rm{cm}})^2(q) \simeq 1 + \sum_{n = 1}^{5} a_n \times 
\begin{cases}
q^{-(n - n_0)} & q \geq 1 \\
q^{(n - n_0)} & q < 1,
\end{cases}
\end{equation}
where $n_0 = 0.658$, $a_{1} = 1.544$, $a_{2} = -3.118$, $a_3 = 4.387$, $a_4 = -3.567$, $a_5 = 1.190$.

For the AML induced by MB,  we adopt the standard Skumanich prescription \citep{1972ApJ...171..565S,1983ApJ...275..713R} with $\gamma_{\rm MB} = 3$. It is worth noting that the effectiveness of MB depends on the mass of the donor's convective envelope. Following \citet{2002ApJ...565.1107P}, for convective envelopes with fractional masses $q_{\rm{conv}}<0.02$, we reduce the MB efficiency by a factor
$e^{1-0.02/q_{\rm{conv}}}$.
 
GW radiation plays a crucial role in short-period binary star systems, and we compute the AML caused by GW using the formula of \citet{1975ctf..book.....L}.

\section{results}
\label{results}

\subsection{The effect of enhanced AML}
\label{The Effect of Mass Loss from $L_2$}

\begin{figure*}
  \begin{minipage}[t]{1\linewidth}
  \centering
  \includegraphics[scale=0.5]{ 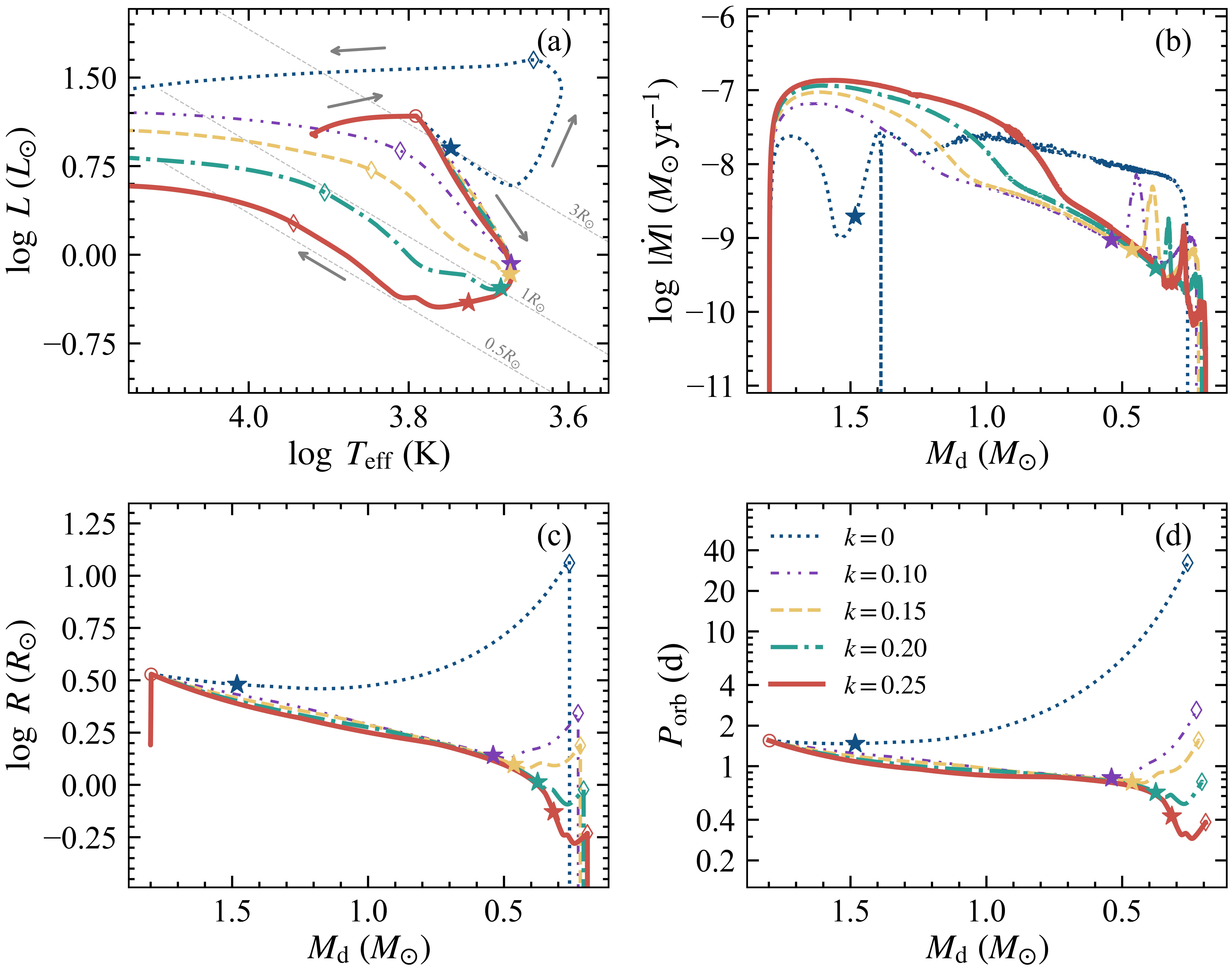}
  \end{minipage}%
  \caption{\label{fig2} The examples of binary evolution with $k$ values of 0 (blue dotted lines), 0.10 (purple dash-dot-dot lines), 0.15 (yellow dashed lines), 0.20 (green dash-dotted lines), and 0.25 (red solid lines) are shown. The initial parameters are the same for all cases: the initial donor mass is $M_{\rm d,i} = 1.8\ M_{\odot}$, the CO WD mass is $M_{\rm CO} = 1.1\ M_{\odot}$, and the initial orbital period $P_{\rm orb, i}$ = 1.55 days.} Panel (a) displays the HR diagram of the donor, with gray dashed lines representing lines of constant radius. Panels (b), (c), and (d) display the mass transfer rate, stellar radius, and orbital period as functions of the donor mass (which decreases from left to right). Circles and diamonds indicate the beginning and end of RLOF, respectively. Stars denote the onset of the He core formation.
\end{figure*}

\begin{figure*}
  \begin{minipage}[t]{1\linewidth}
  \centering
   \includegraphics[scale=0.5]{ 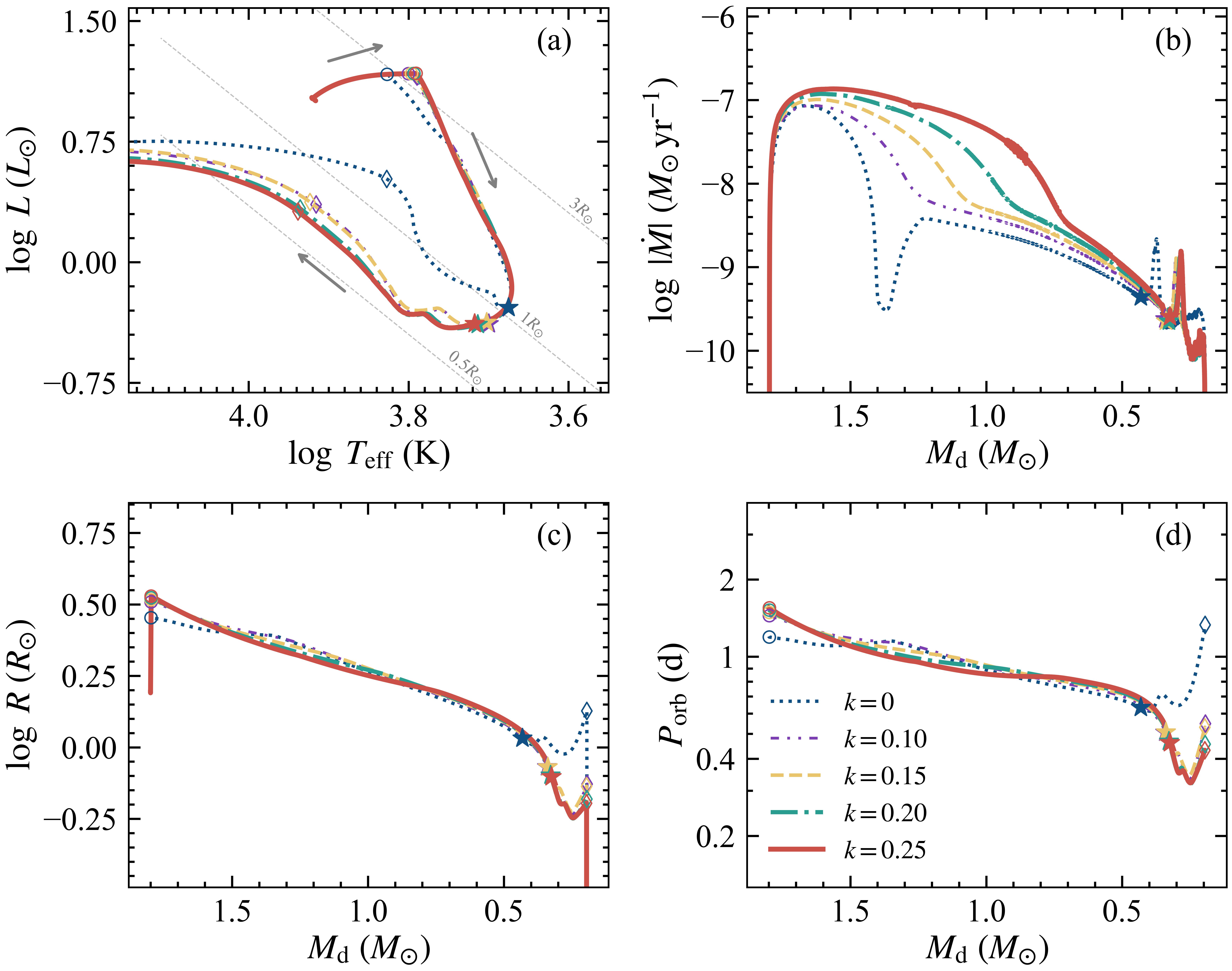}
  \end{minipage}%
  \caption{\label{fig3} Similar to Fig.~\ref{fig2}, but showing examples of binary evolution with the same final ELM WD mass. 
  All systems are calculated with $M_{\rm d,i}=1.8\ M_{\odot}$ and $M_{\rm CO}=1.1\ M_{\odot}$. 
  To produce the same final ELM WD mass, different values of $k$ require different initial orbital periods, 
  namely $k=0$ ($P_{\rm orb,i}=1.201\ \mathrm{d}$), $k=0.10$ ($P_{\rm orb,i}=1.450\ \mathrm{d}$), 
  $k=0.15$ ($P_{\rm orb,i}=1.492\ \mathrm{d}$), $k=0.20$ ($P_{\rm orb,i}=1.523\ \mathrm{d}$), 
  and $k=0.25$ ($P_{\rm orb,i}=1.554\ \mathrm{d}$). 
  The horizontal axis ($M_{\rm d}$) decreases from left to right.}
\end{figure*}

\begin{figure}
  \begin{minipage}[t]{1\linewidth}
  \centering
   \includegraphics[scale=0.52]{ 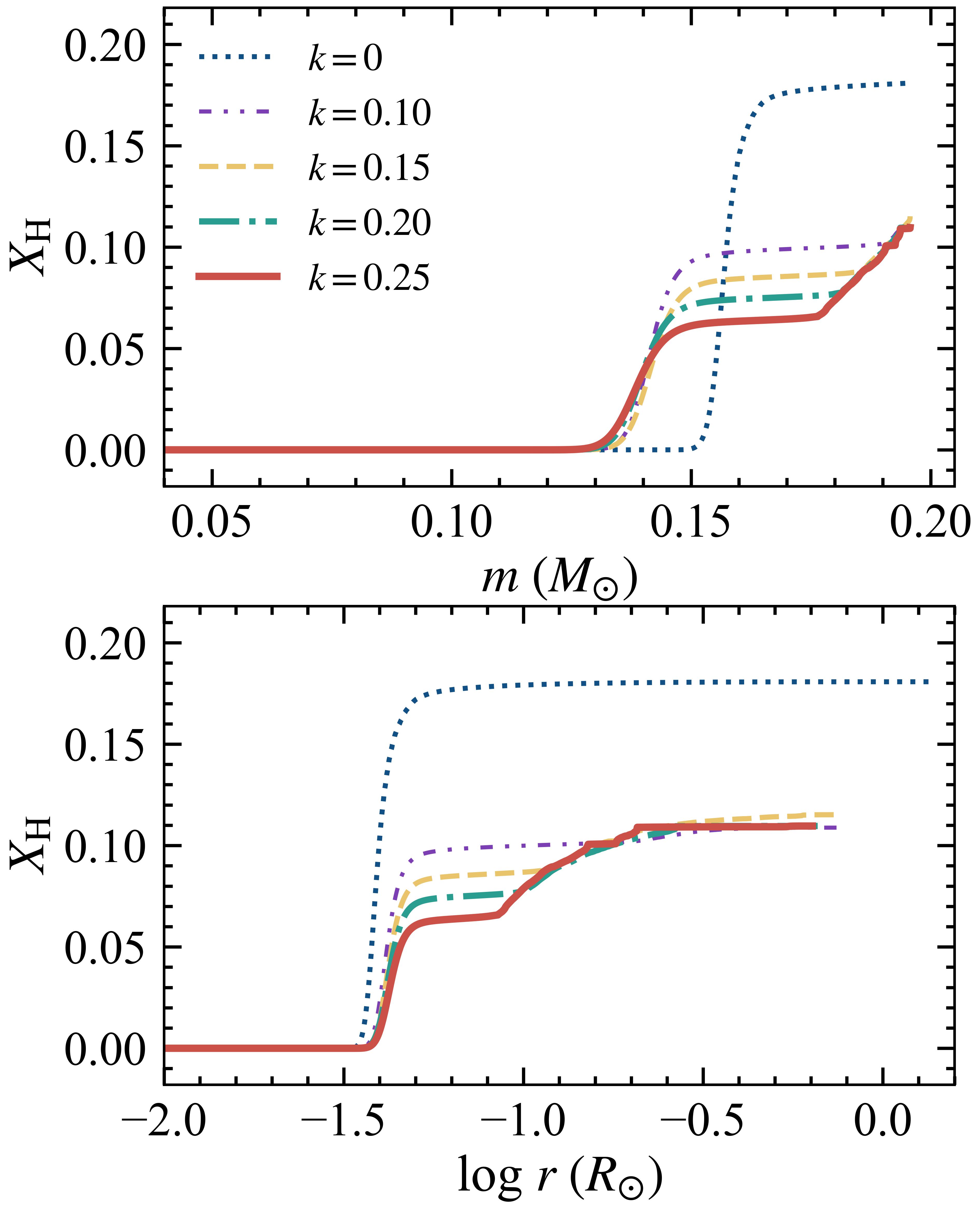}
  \end{minipage}%
  \caption{\label{fig4} The upper and lower panels display the H abundance as functions of donor mass coordinate and radius coordinate at the end of RLOF.} The initial parameters are the same as in Fig.~\ref{fig3}. Different line types represent different k values, as shown in the legend.
\end{figure}
To illustrate the effects of enhanced AML on the binary evolution, 
we present several typical evolutionary tracks in Figure~\ref{fig2}, all with the same initial parameters but different $k$ values: $M_{\rm d, i}=1.8\ M_{\odot}$, $M_{\rm CO}=1.1\ M_{\odot}$, $P_{\rm orb, i}=1.55 \ d$ and $Z=0.02$. Note the horizontal axis decreases from left to right.
The four panels show (a) the evolutionary track of the donor on the Hertzsprung–Russell (HR) diagram, and the relations between the donor mass and (b) the mass transfer rate, (c) the stellar radius, and (d) the orbital period. The star indicates the onset of the He core formation. A larger $k$ value means there is more material lost via the outer Lagrangian point. 

In the examples shown in Figure~\ref{fig2}, the donors fill their Roche lobes at the MS stage and have not yet developed a He core.
As mass transfer proceeds, H fusion continues in the center during mass transfer,
and a He core begins to develop after the donor has lost some material,
as indicated by the stars in Figure~\ref{fig2}.
We note that the onset of He core formation depends on the value of $k$. In general, systems with larger $k$ values form He cores at a later mass transfer stage. This is because the donor is driven out of thermal equilibrium at the onset of mass transfer, and mass transfer is dominated by the thermal expansion,
which is commonly termed thermal-timescale mass transfer \citep[e.g.,][]{1987ApJ...318..794H,2001MNRAS.321..327K,2007A&A...476.1283X,2023pbse.book.....T}.
The stellar thermal timescale is much shorter than the nuclear timescale, with
$t_{\rm th}/t_{\rm nuc} \sim 10^{-2}$--$10^{-3}$ \citep[e.g.,][]{2022abn..book.....C,2023pbse.book.....T,2024MNRAS.531L..45Z}, and thus H burning is inefficient at producing a He core during this phase.
For larger values of $k$, enhanced AML increases the Roche lobe filling factor of the donor, leading to higher mass transfer rates and increased mass loss. As a result, the formation of the He core is delayed to a later stage of mass transfer.

After the formation of the He core, H shell burning drives the radial expansion of the donor, leading to an increase in the mass transfer rate, which corresponds to the peak after He core formation in panel (b) of Figure~\ref{fig2}. For larger values of $k$, the donor retains only a thin H envelope, which is predominantly radiative. The subsequent loss of this H envelope causes further contraction of the donor, thereby reducing the mass transfer rate.
In contrast, in the case of $k = 0$, a substantial H envelope remains at the onset of He core formation. H shell burning drives the donor to ascend the giant branch, during which most of the envelope becomes convective. In this phase, mass loss induces further expansion of the donor, allowing the remaining envelope to be removed at a high mass transfer rate.
The mass transfer phase temporarily halts when the donor mass decreases to approximately $1.4~M_\odot$ (the blue dotted line in panel (b)). During the H shell-burning phase, the nuclear burning rate decreases, leading to a sudden contraction of the stellar radius. This contraction temporarily halts mass transfer, causing the donor to detach from its Roche lobe. Note that MB is negligible in this phase, with a torque of only $\sim 10^{30}$–$10^{33}\ \mathrm{g~cm^2~s^{-2}}$, far smaller than the AML rate due to mass loss \citep[$>10^{35}\ \mathrm{g~cm^2~s^{-2}}$;][]{2002ApJ...565.1107P}.

The radius evolution is shown in panel (c) of Figure~\ref{fig2}. 
A larger value of $k$ leads to a more pronounced reduction in the donor radius after the onset of mass transfer.
This behavior reflects the competition between the donor’s intrinsic thermal response and the orbital evolution driven by AML. 
For systems with small $k$, AML is inefficient and the mass transfer is primarily governed by the donor’s thermal expansion. The orbit gradually widens during the mass transfer phase, accompanied by an expansion of the Roche lobe and an increase in the donor radius \citep{2023hxga.book..129B}.
In contrast, AML dominates the binary evolution for systems with large $k$ (e.g., $k \gtrsim 0.10$). Enhanced AML drives orbital shrinkage, causing the Roche lobe to contract \citep{2023hxga.book..129B}. Consequently, the donor continues to contract during the mass transfer phase until thermal equilibrium is restored. The examples in Figure~\ref{fig2} suggest that enhanced AML would lead to smaller ELM WD mass and shorter orbital period.

We then illustrate how AML aﬀects the WD mass-orbital period relation. In Figure~\ref{fig3}, we show several representative evolutionary tracks with different values of $k$ that produce ELM WDs with the same final mass.
If the initial binary parameters are ﬁxed, a larger value of $k$ leads to a smaller He WD core mass.
This is because enhanced AML drives a stronger mass loss rate.
The rapid reduction in donor mass results in a lower nuclear burning rate in the core, thereby limiting the growth of the He core and producing a smaller final He core mass.
Therefore, to produce ELM WD with equal mass,
mass transfer in binaries with larger $k$ values must be initiated when the donor is at a more advanced evolutionary stage, i.e., at a longer initial orbital period.
From the panel (c) of Figure~\ref{fig3}, we find that a higher value of $k$ results in a smaller donor radius (or equivalently, a shorter orbital period; see panel (d)), even for the same final donor mass. 
This indicates that the strength of AML not only affects the orbital evolution but also leads to different internal structures in donors of the same final ELM WD mass.

The differences in the He WD structures can be seen in Figure~\ref{fig4}, which shows the H abundance profiles of the donors at the end of RLOF for different values of $k$. The He abundance is calculated as $X_{\rm He} = 0.98 - X_{\rm H}$.
We see that the He core mass decreases with increasing $k$.
For example, the system with $k = 0$ develops a He core of $\simeq 0.15\ M_{\odot}$, whereas for $k = 0.10$ the core mass is reduced to $\simeq 0.13\ M_{\odot}$, as shown in the upper panel.
This difference arises because more efficient AML (i.e., $k \gtrsim 0.10$) leads to higher mass loss rates from the donor, which reduce the nuclear reaction rate in the core and result in a smaller core mass for a given final WD mass, as discussed above.
In addition, in systems with larger $k$,
the donor is at a more advanced evolutionary stage when mass transfer begins, resulting in a He-enriched and H-depleted envelope. At the end of mass transfer, the cores are highly degenerate, with only small differences in degeneracy. The core radius therefore has little impact on the stellar radius, which is largely determined by the envelope structure, as shown in the lower panel.
The donor therefore becomes more compact, and systems with larger $k$ produce donors with smaller radius, consistent with the radius differences shown in panel (c) of Figure~\ref{fig3}.

\subsection{The $\log T_{\rm eff}-\log g$ plane}
\label{The surface gravity-effective temperature plane}

\begin{figure}
  \begin{minipage}[t]{1\linewidth}
  \centering
   \includegraphics[scale=0.45]{ 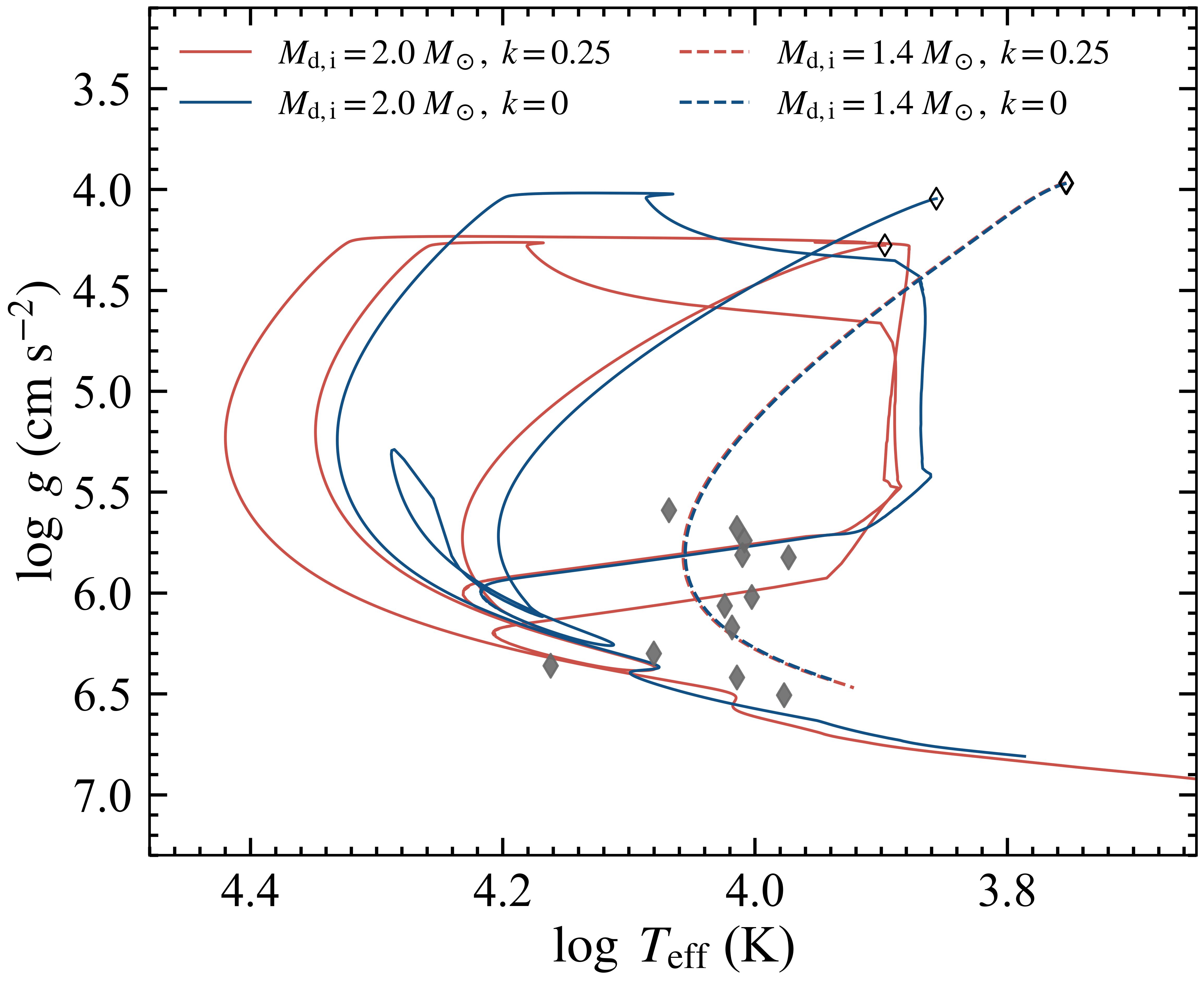}
  \end{minipage}%
  \caption{\label{fig9} Evolutionary tracks in the $\log T_{\rm eff}-\log g$ plane for a $\sim0.168\ M_\odot$ He WD produced from different donor masses with $Z=0.02$, covering the evolution from Roche-lobe detachment to 13.7 Gyr. Solid lines represent progenitor masses of $ 2.0 \ M_{\odot} $, whereas dashed lines denote progenitor masses of $ 1.4 \ M_{\odot} $. The black open diamonds indicate the end of RLOF. Red and blue lines correspond to $k=0.25$ and $k=0$, respectively. Gray filled diamonds denote observed ELM WDs of $0.168 \pm 0.01\ M_\odot$ formed via the RLOF channel.}
\end{figure}

\begin{figure}
  \begin{minipage}[t]{1\linewidth}
  \centering
   \includegraphics[scale=0.38]{ 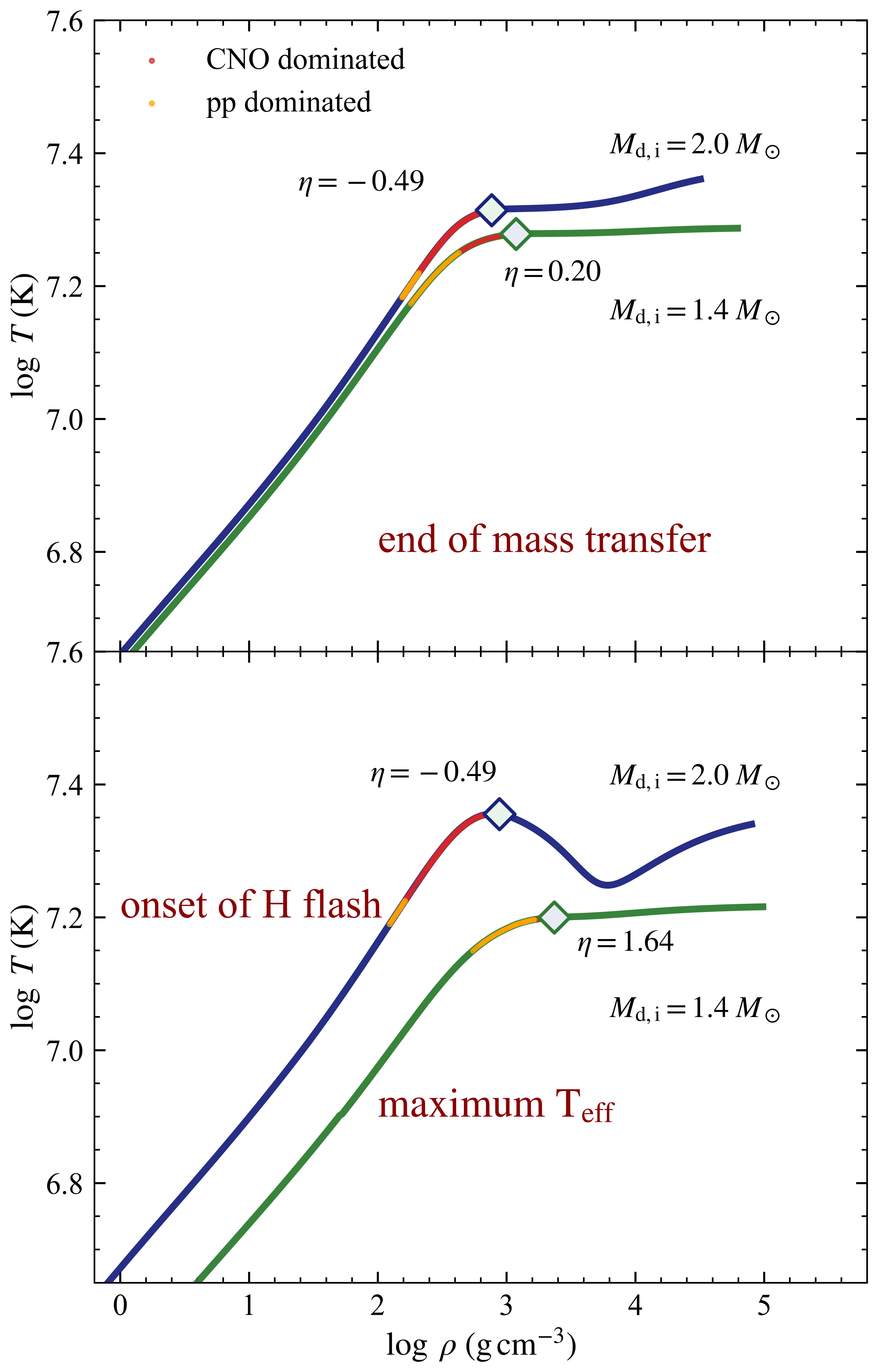}
  \end{minipage}%
  \caption{\label{fig13} Temperature–density profiles of $2.0\ M_{\odot}$ (dark blue line) and $1.4\ M_{\odot}$ (green line) WD progenitors with $k=0$. The upper panel shows models at the end of mass transfer, while the lower panel shows the same $2.0\ M_{\odot}$ progenitor at the onset of the H flash and the $1.4\ M_{\odot}$ model entering the cooling phase. Diamond symbols mark the core–envelope transition, where the electron degeneracy parameter $\eta$ is the value at the core boundary. Red and yellow regions mark where the nuclear energy generation rate exceeds $1\ \rm{erg \ g^{-1}}\ s^{-1}$, with red denoting CNO-dominated and yellow pp-dominated.}
\end{figure}

We then investigate the influence of AML on the evolutionary tracks of ELM WDs in the $\log T_{\mathrm{eff}} - \log g $ plane. Figure~\ref{fig9} shows a set of calculated evolutionary tracks for ELM WDs with a fixed mass of $0.168 \ M_{\odot} $. Solid lines represent progenitor masses of $ 2.0 \ M_{\odot} $, while dashed lines correspond to progenitors of $1.4 \ M_{\odot}$. The red and blue curves denote models with $ k = 0.25 $ and $k = 0$, respectively. The tracks span from the point of Roche-lobe detachment (marked by open diamonds) up to an age of 13.7 Gyr. It is evident that for binaries with the same progenitor, varying the parameter $k$ has only a minor influence on the evolutionary tracks. In particular, for low-mass progenitors, the He core developed during the mass transfer phase becomes fully degenerate. Consequently, the manner of AML has almost no discernible effect on the resulting evolutionary tracks.

The progenitor mass, on the other hand, significantly affects the evolutionary track of an ELM WD of the same mass. As seen in Figure~\ref{fig9}, ELM WDs originating from a $2.0\ M_{\odot}$ progenitor have higher effective temperatures and larger surface gravities at the end of mass transfer than those from a $1.4\ M_{\odot}$ progenitor. This difference arises primarily from the distinct envelope structure of the resulting (pre-)ELM WD. In a more massive progenitor, more H is fused in the envelope, producing a He-enriched envelope. Consequently, the (pre-)ELM WD has a smaller radius, which leads to a higher effective temperature and larger surface gravity.

Moreover, the (pre-)ELM WD from a $2.0\ M_{\odot}$ progenitor is more likely to undergo an H-shell flash. The ignition of an H-shell flash is sensitive to the shell structure; generally, shell flashes develop in a layer where the degeneracy parameter ($\eta = \mu_{\rm electron}/k_{\rm B}T_{\rm B}$) is close to zero \citep{2004ApJ...616.1124N}. Figure~\ref{fig13} shows the temperature–density profiles for the $2.0\ M_{\odot}$ progenitor and the $1.4\ M_{\odot}$ model. The upper panel corresponds to the stage just after mass transfer ends, while the lower panel shows the profiles of the $2.0\ M_{\odot}$ progenitor at the onset of the H flash, together with the $1.4\ M_{\odot}$ model as it enters the cooling phase (i.e., at the maximum effective temperature). At similar evolutionary stages, the density at the base of the H envelope in the $2.0\ M_{\odot}$ model is slightly lower than in the $1.4\ M_{\odot}$ model. However, the corresponding temperature is significantly higher, making the region conducive to unstable CNO burning, which triggers the H flash. Although the envelope is more He-rich, the H mass fraction at the burning layer remains sufficient to sustain CNO burning. In contrast, in the $1.4\ M_{\odot}$ model, the temperature at the base of the H envelope is lower, preventing efficient CNO burning even though the region is more degenerate.

These flashes further reduce the H envelope mass, leading to shorter cooling timescales, which in turn further reduce the radius and increase $\log g$ during the ELM WD cooling stage \citep{2000MNRAS.316...84S,2014A&A...571L...3I,2016A&A...595A..35I}.

\subsection{The ELM WD mass-orbital period plane}
\label{The ELM WD mass-orbital period plane}

\begin{figure}
  \begin{minipage}[t]{1\linewidth}
  \centering
   \includegraphics[scale=0.48]{ 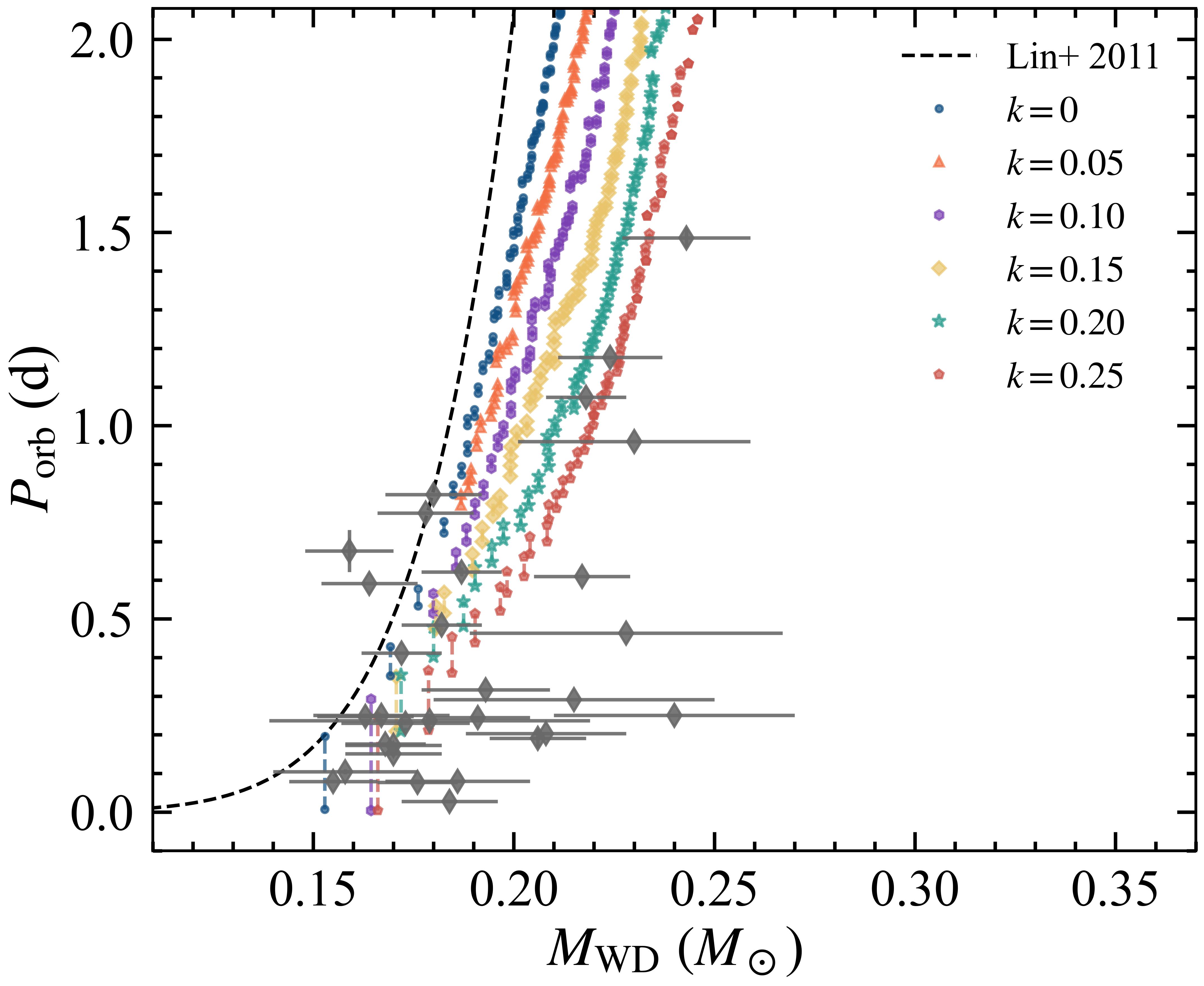}
  \end{minipage}%
  \caption{\label{fig5} The donor mass-orbital period panel for different values of $k$, with $M_{\rm d,i} = 2.0\ M_{\odot}$ and $M_{\rm CO} = 0.8\ M_{\odot}$, under different initial orbital periods. Gray diamonds denote ELM WDs formed via the RLOF channel. The black dashed line is taken from \citet{2011ApJ...732...70L} based on detailed binary evolution calculations.}
\end{figure}

\begin{figure}
  \begin{minipage}[t]{1\linewidth}
  \centering
   \includegraphics[scale=0.48]{ 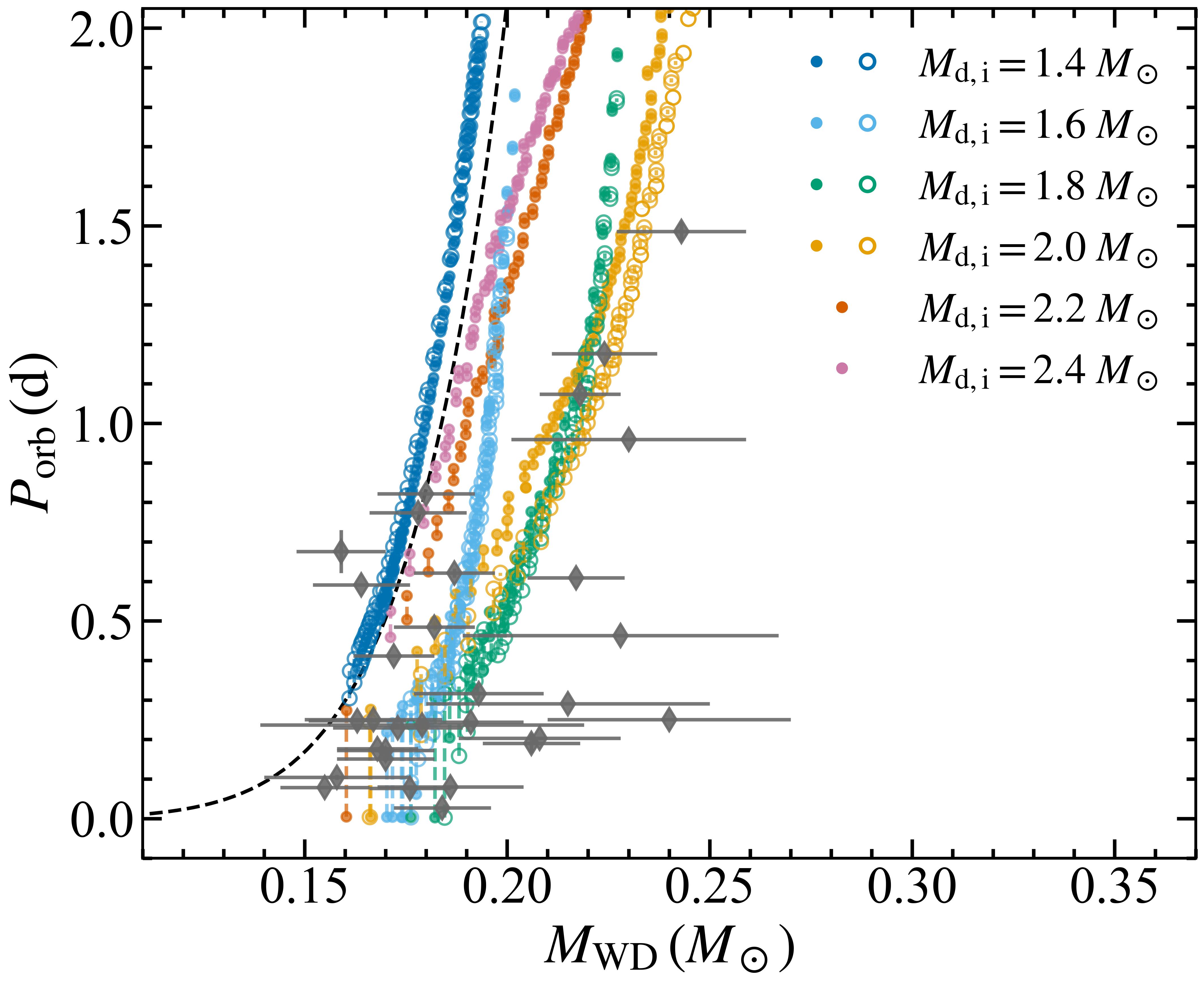}
  \end{minipage}%
  \caption{\label{fig6} Similar to Figure~\ref{fig5}, but for $k$ = 0.25 and donor initial masses ranging from 1.4 to 2.4 $M_\odot$. Open and filled circles denote systems with accretor initial masses of 0.8 and 1.1 $M_\odot$, respectively.}
\end{figure}

We computed a series of binaries with different values of $k$,
and compared the simulated results with observations on the WD mass–orbital period plane with $M_{\rm d, i}=2.0\ M_{\odot}$ and  $M_{\rm CO}=0.8\ M_{\odot}$.
Figure~\ref{fig5} shows the resulting relation between the He WD mass and the orbital period for different $k$ values.
The black dashed line represents the theoretical WD mass–orbital period relation for donors with degenerate cores.
Grey diamonds denote the observed ELM WDs formed through the RLOF channel, as classified based on the results shown in Figure~\ref{fig1}.
The orbital period evolution due to GW radiation is indicated by the colored dashed lines connecting two same symbols.
We found that for a given He WD mass, a larger $k$ value would give a more compact He WD and lead to a shorter orbital period, as discussed in Section~\ref{The Effect of Mass Loss from $L_2$}.

The inner structure is different for different initial donor masses, so the initial donor mass has an eﬀect on the $M_{\rm WD}-P_{\rm orb}$ relation. In Figure~\ref{fig6}, we present the $M_{\rm WD}-P_{\rm orb}$ panel for donors of different initial masses
with accretors set at two representative masses of 0.8 and 1.1 $M_\odot$, for clarity, only the case of $k=0.25$ is shown. The $M_{\rm WD}-P_{\rm orb}$ relation is primarily determined by the donor mass and initial orbital period, while its dependence on the accretor mass is weak.
The core is degenerate if the donor mass is \(1.4~M_\odot\) \citep{2019ApJ...871..148L}, and the final WD mass is largely determined by the core mass, which leads to an \(M_{\rm WD}\)–\(P_{\rm orb}\) relation consistent with the results of \citet{2011ApJ...732...70L}.

Our results indicate that simulations including enhanced AML can reproduce the majority of the observed systems. Only a few systems with relatively massive He WDs are not reproduced, which may be attributed to differences in metallicity, as discussed below.

\subsection{The effect of metallicity }
\label{The effect of metallicity}

\begin{figure*}
  \begin{minipage}[t]{1\linewidth}
  \centering
   \includegraphics[scale=0.56]{ 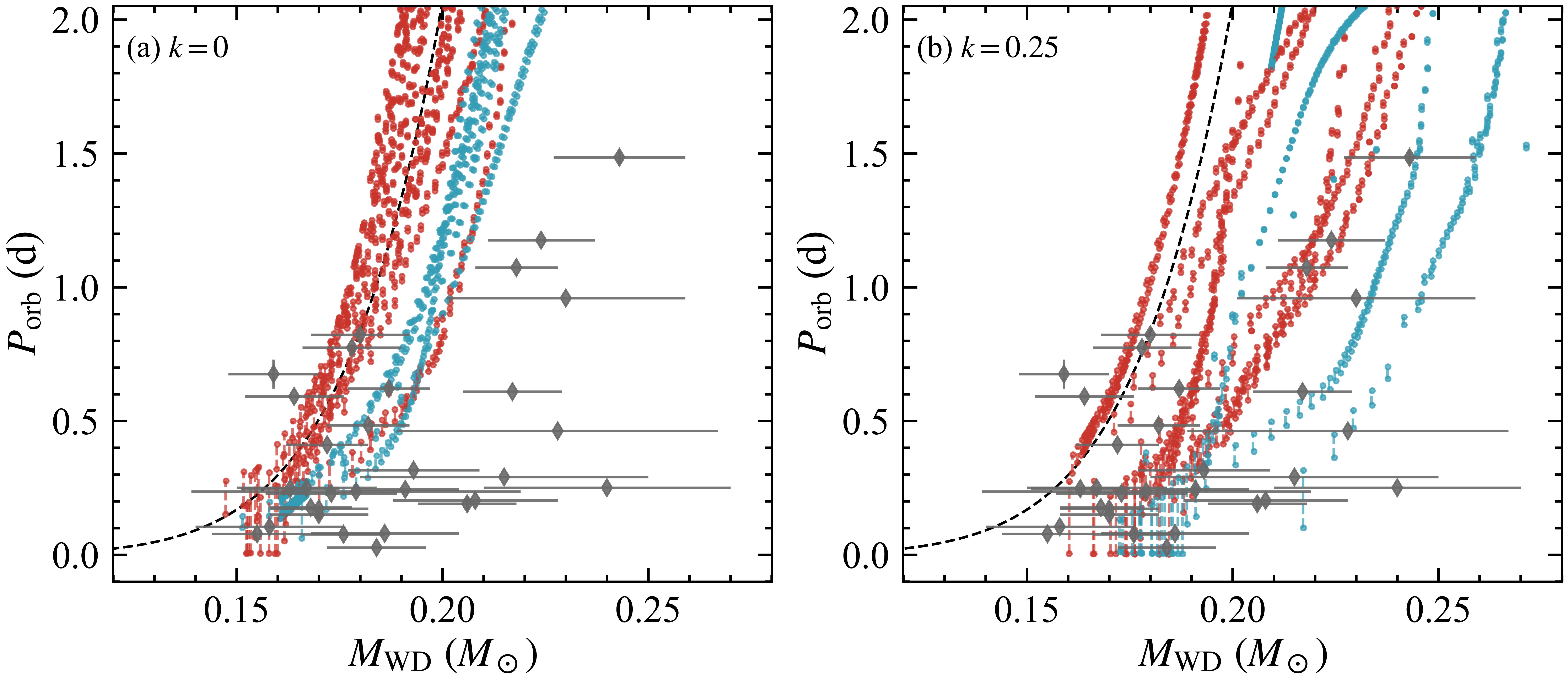}
  \end{minipage}%
  \caption{\label{fig7} The $M_{\rm WD}$–$P_{\rm orb}$ panel with $k = 0$ and $k = 0.25$. The red and blue solid circles represent systems with solar metallicity ($Z = 0.02$) and low metallicity ($Z = 0.001$), respectively. The initial binary mass grids are: $k=0$, $Z=0.02$: $(M_{\rm CO}, M_{\rm d,i}) = (1.1, 1.4–2.8)$ and $(0.8, 1.4–2.4) \ M_\odot$; $Z=0.001$: $(1.1, 1.4–2.4)$ and $(0.8, 1.4–2.2) \ M_\odot$. $k=0.25$, $Z=0.02$: $(1.1, 1.4–2.4)$ and $(0.8, 1.4–2.0) \ M_\odot$; $Z=0.001$: $(1.1, 1.4–2.0)$ and $(0.8, 1.4–1.6) \ M_\odot$. }
\end{figure*}

Metallicity is an important parameter that affects the evolutionary track of stars and binaries \citep[e.g.,][]{2010ApJ...715L.138B,2017A&A...597A..71S,2020A&A...638A..55K}.
Meanwhile, observations show that ELM WDs exist in environments spanning the Galactic disk to halo, indicating that they can originate from different metallicities
\citep[e.g.,][]{2015MNRAS.453.2707R, 2015ApJ...812...63C},
and we compute a set of binaries by considering a low metallicity of $Z = 0.001$.
Figure~\ref{fig7} shows the resulting $M_{\rm WD}$–$P_{\rm orb}$ relations for $k = 0$ (left panel) and $k = 0.25$ (right panel) with different metallicities. 

From the panel (a) of Figure~\ref{fig7}, it is clear that lower metallicity corresponds to shorter orbital periods. Low-metallicity stars are generally hotter and more compact owing to their lower radiative opacity \citep{1983A&A...119...54E,2001A&A...371..152M}. As a result, RLOF occurs at smaller orbital separations, leading to systematically shorter final orbital periods for a given WD mass. However, in the case of standard AML, considering low metallicity alone still cannot account for the existence of observations with relatively massive ELM WDs and short orbital periods (see also \citealt{2019ApJ...871..148L}). In panel (b), we present the enhanced AML with $k = 0.25$, it gives the relatively shorter final orbital periods as expected.
The observed orbital period distribution of ELM WD binaries can be well reproduced when both metallicity effects and enhanced AML are taken into account, indicating that our results are applicable to both the disk and halo populations of ELM WDs.

It should be noted that the adopted value of $k = 0.25$ in this work represents an upper limit for the magnitude of AML. In reality, the value of $k$ may vary across different binary systems. Generally, a larger $k$ leads to a shorter orbital period, while systems with long orbital periods likely correspond to a lower $k$. Furthermore, although $k$ is treated as a constant during the mass transfer phase in this study, it may in fact vary depending on factors such as the mass transfer rate and orbital parameters \citep{2023MNRAS.519.1409L,2024ApJ...977...34T}. Consequently, our results establish a lower limit on the orbital period under the condition of enhanced AML.

\section{Discussions: Mass loss versus MB}
\label{Mass loss versus MB}
In this work, we investigate the extra AML case where additional angular momentum is carried away by material lost through the $L_2$ point. In fact, other mechanisms may also contribute to AML during mass transfer. One of the most well-studied processes is enhanced MB \citep{2024A&A...682A..33B,2024PASP..136l4201B,2025A&A...696A..92B}. Some studies have suggested that the widely adopted Skumanich MB prescription faces difficulties in explaining the observed populations of low-mass X-ray binaries (LMXBs) or AM CVn systems \citep{2019MNRAS.483.5595V,2023A&A...678A..34B}. Enhanced MB models assume that more angular momentum is removed by magnetized stellar winds, which would have analogous effects on orbital evolution as mass loss through the $L_2$ point. Here, we adopt the Convection And Rotation Boosted (CARB) MB model \citep{2019ApJ...886L..31V} as an example to investigate its impact on orbital evolution and to compare with our extra AML results.

The CARB MB model incorporates two recent improvements in the understanding of stellar magnetic fields and magnetized winds: the dependence of magnetic field strength on the depth of the outer convective zone, and the dependence of the Alfven radius on the donor's rotation. This model provides a better match to the observed persistent neutron star LMXBs and ultra-compact X-ray binaries \citep[UCXBs; see][]{2019ApJ...886L..31V,2021ApJ...909..174D, 2021MNRAS.503.3540C}.

\begin{figure}
  \begin{minipage}[t]{1\linewidth}
  \centering
   \includegraphics[scale=0.41]{ 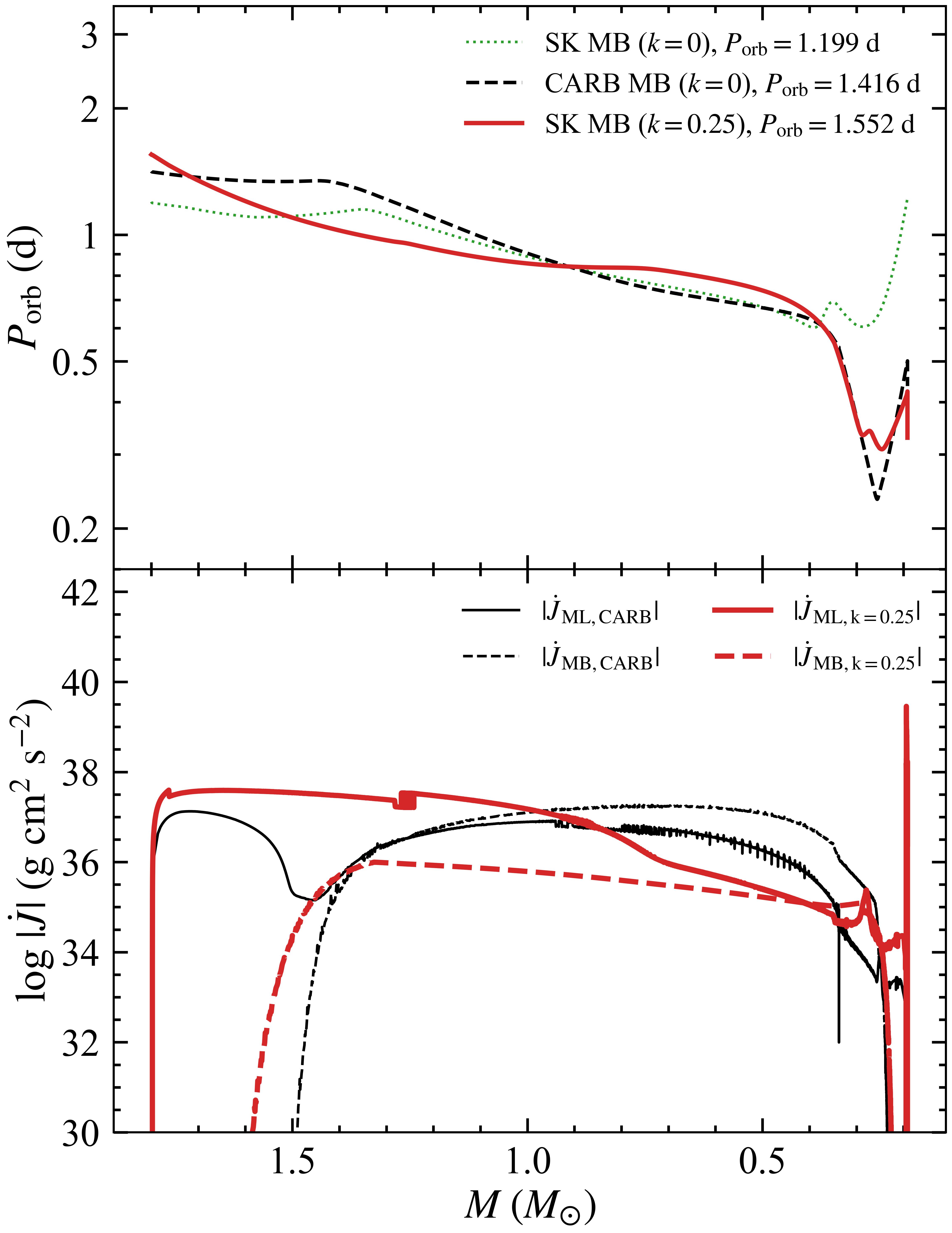}
  \end{minipage}%
  \caption{\label{fig11} Upper panel: Comparison of the $M_{\rm d}$–$P_{\rm orb}$ plane for three models, SK MB with $k=0$, CARB MB with $k=0$, and SK MB with $k=0.25$. All models produce the same final ELM WD mass of $\sim 0.192\ M_{\odot}$, with initial parameters $M_{\rm d,i}=1.8\ M_{\odot}$ and $M_{\rm CO,i}=1.1\ M_{\odot}$. Lower panel: The AML rates due to MB ($\dot{J}_{\rm MB}$) and mass loss ($\dot{J}_{\rm ML}$) as a function of the donor mass, comparing CARB MB with $k=0$ ($\dot{J}_{\rm MB, CARB}$, $\dot{J}_{\rm ML, CARB}$) and SK MB with $k=0.25$ ($\dot{J}_{\rm MB, k=0.25}$, $\dot{J}_{\rm ML, k=0.25}$).}
\end{figure}

In the upper panel of Figure~\ref{fig11}, we present the $M_{\rm d}$–$P_{\rm orb}$ plane for an initial system with $M_{\rm d,i}=1.8\ M_\odot$ and $M_{\rm CO,i}=1.1\ M_\odot$. Three models are considered: SK MB with $k=0$, SK MB with $k=0.25$, and CARB MB (standard mass-loss scheme, $k=0$). All three models produce the same final ELM WD mass of $\sim 0.192\ M_\odot$. Compared to the standard MB case ($k=0$, green dotted line), both the CARB MB model and the $k=0.25$ model yield shorter orbital periods at the same final donor mass, as expected, because both prescriptions remove more orbital angular momentum during the mass transfer phase.

We further compare the difference between the CARB MB model and the $k=0.25$ case. In the lower panel of Figure~\ref{fig11}, we show the evolution of the AML rates due to MB and mass loss as a function of the donor mass. At the beginning of mass transfer, the donor star has not yet developed a convective envelope, so MB is inactive. The dominant AML mechanism is mass loss, with the $k=0.25$ case removing more angular momentum via mass loss than the $k=0$ case. As the donor mass decreases to about $1.5\ M_\odot$, the star gradually develops a substantial convective envelope, and MB becomes effective. In the SK MB model (red dashed line in the lower panel), the AML rate is about $10^{35-36}\ \rm g~cm^2~s^{-2}$, whereas the CARB MB model yields a significantly higher AML rate, with an average value of $\sim 10^{37}\ \rm g~cm^2~s^{-2}$, which is one to two orders of magnitude higher than that of the SK MB model. Overall, the $k=0.25$ case enhances AML during the early stage of mass transfer, while the CARB MB model enhances AML during the later stage. Because the peak AML rates of $\dot{J}_{\rm ML,k=0.25}$ and $\dot{J}_{\rm MB,CARB}$ are quite close, the resulting final orbital periods for these two cases are similar. Therefore, enhanced MB may serve as an alternative pathway to achieve extra AML. However, it should be noted that MB is driven by stellar winds, and the AML caused by MB should be positively correlated with the wind strength \citep[as described in][]{2019MNRAS.483.5595V}. For low-metallicity stars (such as those in ELM WD binaries found in the halo), the AML due to MB should be suppressed. In contrast, our extra AML model via mass loss is almost unaffected by metallicity. This allows our results to reproduce most of the observed samples in the WD mass–orbital period plane (Figure~\ref{fig7}).

\begin{figure}
  \begin{minipage}[t]{1\linewidth}
  \centering
   \includegraphics[scale=0.34]{ 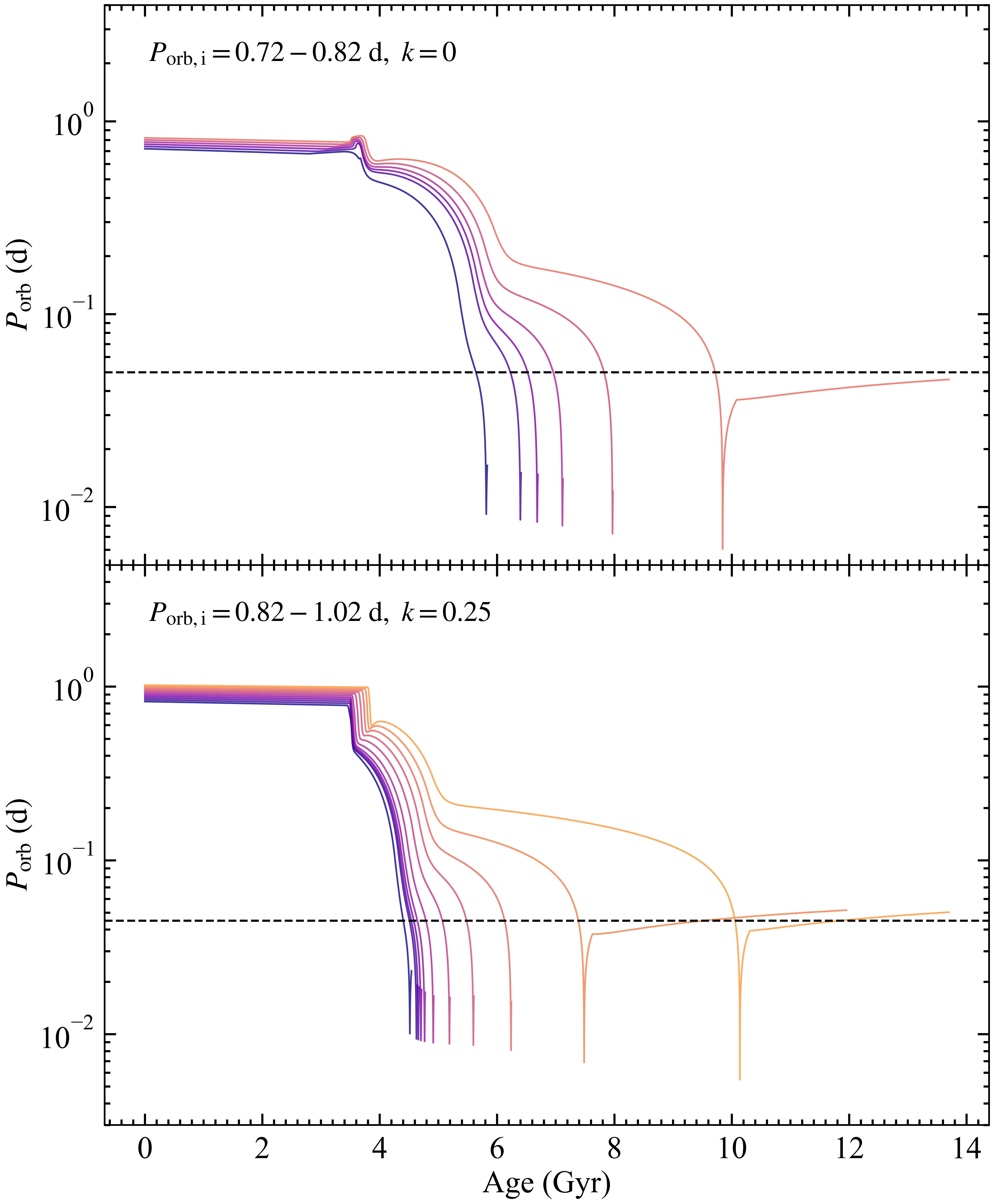}
  \end{minipage}%
  \caption{\label{fig12} Evolution of the orbital period as a function of time for systems with $M_{\rm d,i}=1.3\ M_{\odot}$ and $M_{\rm CO,i}=1.1\ M_{\odot}$, and different initial orbital periods in steps of 0.02 days. The upper panel corresponds to $k=0$, and the lower panel corresponds to $k=0.25$. The black dashed line indicates an orbital period of $\sim 65$ min.}
\end{figure}

Some recent works have proposed that enhanced MB models can expand the parameter space for LMXBs/UCXBs \citep{2021ApJ...909..174D, 2021MNRAS.503.3540C}. Here we show that extra AML caused by mass loss exhibits similar behaviour. In Figure~\ref{fig12}, we compare the parameter space for AM CVn formation under the standard AML prescription (upper panel, $k=0$) and the enhanced AML case (lower panel, $k=0.25$), with initial parameters $M_{\rm d,i}=1.3\ M_\odot$ and $M_{\rm CO,i}=1.1\ M_\odot$. We set an orbital period of $\sim 65$ min as a practical threshold for identifying AM CVn systems, which is consistent with observations \citep{2010PASP..122.1133S}, as indicated by the black dashed line. For $k=0$, the initial orbital period range for forming AM CVn systems is $0.72$–$0.82$ days, while for $k=0.25$ it expands to $0.82$–$1.02$ days. Thus, the allowable parameter space is increased by a factor of two. These results suggest that extra AML caused by mass loss may be an alternative route to explain the observed populations of AM CVn or UCXB systems.

\section{Summary and Conclusion}
\label{conclusion}

In this work, we investigate the effect of enhanced AML on the formation of ELM WD binaries in the stable RL channel.
We compute a grid of models for ELM WDs formed through the stable mass transfer channel by considering the enhanced AML.
We parameterize the AML using a dimensionless parameter $k$, which represents the fraction of mass lost through the outer Lagrange point. 
We investigate the impact of AML on the donor’s evolutionary properties and the orbital period, and further examine the effects of the initial binary masses and metallicity on the $M_{\rm WD}$–$P_{\rm orb}$ relation. Our main conclusions are summarized as follows:

\begin{enumerate}
    \item Based on our reclassification of the updated observed samples for ELM WD binaries in the ELM survey, we find that $\sim68\%$ of systems are formed through the RLOF channel instead of the CE channel.    
    \item Enhanced AML drives a higher mass transfer rate,
    thereby causing more rapid mass loss from the donor on a thermal timescale.
    This fast mass loss suppresses the growth of the core, thus resulting in the formation of a smaller He core.
    Moreover,
    stronger AML leads to a more He-enriched (H-depleted) envelope for a given final He WD mass,
    which produces a more compact He WD and a shorter orbital period.
    \item Binaries with low-metallicity donors generally have shorter orbital periods than solar-metallicity systems. Compared with the standard AML mechanisms, our simulations with enhanced AML successfully reproduce the full parameter space covered by the ELM Survey.
\end{enumerate}

Our findings indicate that the standard AML mechanism, where unprocessed material carries away the specific angular momentum of the accretor’s surface, is insufficient to account for the observed orbital period distributions of ELM WD binaries. 
In particular, this conventional prescription fails to reproduce the significant population of systems with periods shorter than those predicted by theoretical models of stable mass transfer. An enhanced AML mechanism offers a promising pathway to resolve this discrepancy, as it would allow binaries to shed additional angular momentum and thus evolve into tighter orbits. 
Nevertheless, significant uncertainties remain in quantifying AML. Several key questions are still unresolved.
First, the magnitude of the extra AML is poorly constrained and likely depends on parameters like mass ratio, orbital period and mass transfer rate.
Second, the effects of other physical parameters on AML remain unclear, especially its dependence on the orbital period and the mass-transfer rate.
Furthermore, it remains unclear whether enhanced AML operates predominantly during active mass transfer phases or is triggered at specific evolutionary stages. 
The need for such a mechanism appears to extend beyond ELM WDs;
populations such as WD + NS binaries and subdwarf B binaries \citep[e.g.,][]{2005AJ....129.1993M,2020NatAs...4...72C,2021MNRAS.507.2137R,2022ApJ...940...86K,2025arXiv251120147M} also exhibit orbital periods shorter than those predicted by canonical stable mass transfer simulations, suggesting that enhanced AML may be a broader and more fundamental phenomenon in close binary evolution.
Consequently, the definitive resolution of these issues awaits future investigations.

\section*{Acknowledgements}
This work is partially supported by the Natural Science Foundation of China (grant Nos.~12125303, 12288102, 12473034, 12273105, 11703081, 11422324, 12073070, 12525304, 12473033, 12333008, 12422305), the Strategic Priority Research Program of the Chinese Academy of Sciences (grant Nos.~XDB1160201, XDB1160303, XDB1160300, XDB1160000), the National Key R\&D Program of China (grant Nos.~2021YFA1600401, 2021YFA1600403, 2021YFA1600400), the Yunnan Revitalization Talent Support Program-Science \& Technology Champion Project (No.~202305AB350003), the Young Talent Project of Yunnan Revitalization Talent Support Program, Yunnan Fundamental Research Projects
(Nos.~202401BC070007, 202401AT070139), the International Centre of Supernovae (ICESUN), Yunnan Key Laboratory of Supernova Research (No. 202505AV340004), the Yunnan Revitalization Talent Support Program ``YunLing Scholar'' project, the New Cornerstone Science Foundation through the XPLORER PRlZE, the Key Research Program of Frontier Sciences of CAS (No. ZDBS-LY-7005).

\software{MESA
\citep[v12115;][]{2011ApJS..192....3P,2013ApJS..208....4P,2015ApJS..220...15P,2018ApJS..234...34P,2019ApJS..243...10P,2023ApJS..265...15J} }

Example MESA inlists used in this work are archived on Zenodo and are available at \url{https://doi.org/10.5281/zenodo.19784240}.
%%%%%%%%%%%%%%%%%%%%%%%%%%%%%%%%%%%%%%%%%%%%%%%%%%
%\section*{Data Availability}

%The data underlying this article will be shared on reasonable request to the corresponding author.

%% For this sample we use BibTeX plus aasjournalv7.bst to generate the

\appendix

\section{Core-Boundary Effects on ELM WD Formation-Channel Classification}
\label{appendix}

\begin{figure*}
  \begin{minipage}[t]{1\linewidth}
  \centering
   \includegraphics[scale=0.56]{ 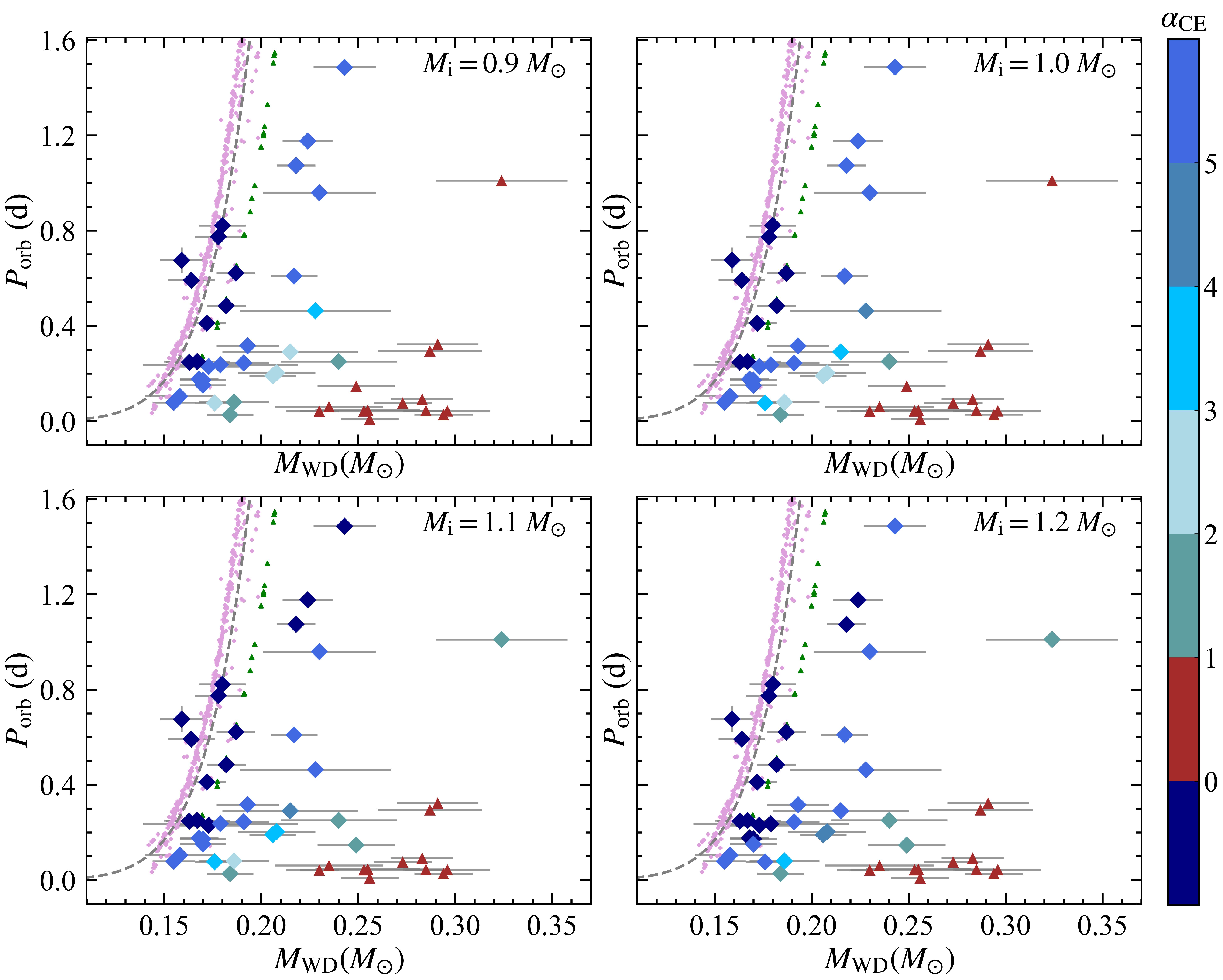}
  \end{minipage}%
  \caption{\label{fig8} Similar to Figure~\ref{fig1}, but the boundary between the core and the envelope is defined at the location where the H mass fraction equals 0.1.
}

\end{figure*}

We investigate the impact of the adopted core-boundary definition on the classification of formation channels for ELM WD binaries.
Figure~\ref{fig8} shows the classification of the observed ELM WD systems when adopting $X_{\rm H}=0.1$ as the boundary between the He core and the H envelope.
We find that a higher $X_{\rm H}$ definition of the core boundary requires a smaller $\alpha$ for the formation of ELM WD binaries.
However, the overall classification of the formation channels is not significantly affected.

\section{The $\gamma$ Mechanism in the Classification of ELM WD Formation Channels}
\label{appendixb}

Here we test whether the $\gamma$ mechanism can account for the observed sample. Following \citet{2000A&A...360.1011N,2005MNRAS.356..753N,2023ApJ...944...87D,2024ApJ...961..202G}, the $\gamma$ mechanism is expressed as:

\begin{equation}
\begin{aligned}
& \frac{a_i}{a_{\mathrm{f}}}=\left[\frac{1}{1+q_{\mathrm{ec}}}\right]^2\left[\frac{1+q\left(1+q_{\mathrm{ec}}\right)}{1+q}\right] \times\left[\frac{1+q\left(1+q_{\mathrm{ec}}\right)}{1+q+q q_{\mathrm{ec}}(1-\gamma)}\right]^2 .
\end{aligned}
\end{equation}
where $q_{\rm ec} = M_{\rm env}/M_{\rm c}$.
We adopt $\gamma=1.5$ as the boundary between RLOF and CE channel \citep{2005MNRAS.356..753N}. Smaller $\gamma$ values imply insufficient AML to reproduce the observations.

\begin{figure*}
  \begin{minipage}[t]{1\linewidth}
  \centering
   \includegraphics[scale=0.46]{ 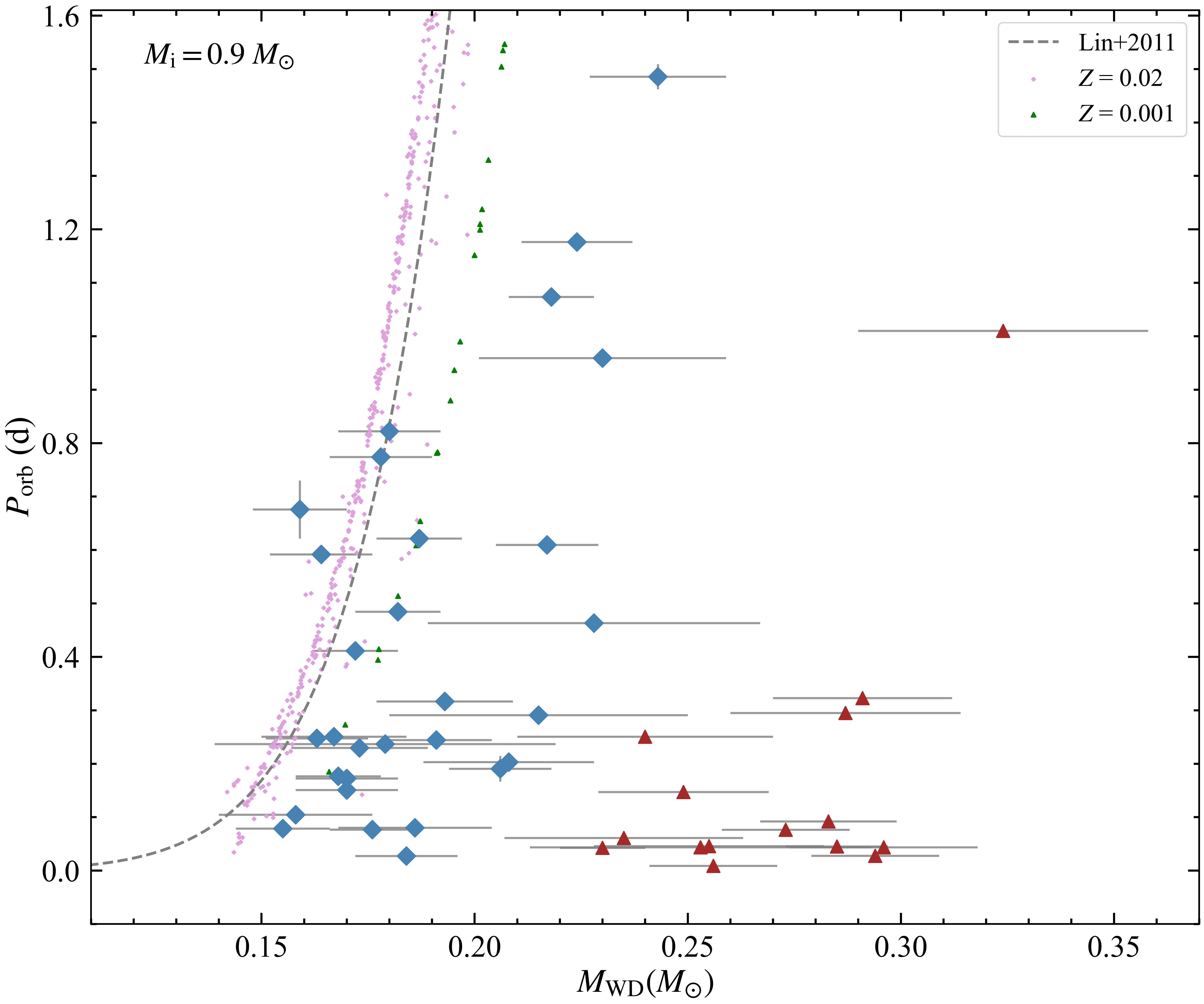}
  \end{minipage}%
  \caption{\label{fig10} 
  The $M_{\rm WD}$–$P_{\rm orb}$ relation for systems with MS progenitors of initial mass $0.9\ M_{\odot}$. Assuming all observed systems formed via CE ejection, we compute the corresponding CE parameter $\gamma$ based on angular momentum balance. Systems with $\gamma < 1.5$ are shown as blue diamonds, while those with $\gamma \geq 1.5$ are shown as red triangles. Other symbols are the same as in Figure~\ref{fig1}.
}

\end{figure*}

Assuming all observed ELM WDs are formed via CE ejection but adopting the $\gamma$ mechanism, we compute the corresponding $\gamma$ parameter for all observational samples. The results are shown in Figure~\ref{fig10}. Samples with $\gamma \ge 1.5$ are regarded as formed via the CE channel and are shown as red triangles, while those with $\gamma < 1.5$ are shown as blue diamonds, indicating formation through the RLOF channel. The results indicate that even with the $\gamma$ formalism, it remains difficult to fully account for the samples with unreasonable $\alpha$ values in the energy balance. Both the $\gamma$ and $\alpha$ mechanisms yield similar observed ELM WD binaries for the RLOF channel and therefore do not affect our subsequent conclusions.

\bibliography{sample701}{}
\bibliographystyle{aasjournalv7}

\end{document}